\documentclass[twocolumn]{aastex6}

\usepackage{graphicx}
\usepackage{epstopdf}
\usepackage{dcolumn}
\usepackage{bm}
\usepackage{color}
\usepackage{epsfig}
\usepackage{amsmath}
\usepackage[caption=false]{subfig}
\usepackage{natbib}

\newcommand\mpo{\textcolor{red}}

\def\ee{\end{equation}}
\def\be{\begin{equation}}
\def\bdm{\begin{displaymath}}
\def\edm{\end{displaymath}}
\def\ebe{\end{displaymath}\begin{equation}}

\def\jn2{J_n^2(z)}

\def\l{\left}
\def\r{\right}

\def\bc{\beta _c}
\def\pa{\partial }

\def\bea{\begin{eqnarray}}
\def\eea{\end{eqnarray}}
\def\bd{\begin{displaymath}}
\def\ed{\end{displaymath}}
\def\ba{\begin{array}}
\def\ea{\end{array}}
\def\ebe{\edm\be}

\def\bea{\begin{eqnarray}}
\def\eea{\end{eqnarray}}
\def\ban{\begin{eqnarray*}}
\def\ean{\end{eqnarray*}}
\def\bd{\begin{displaymath}}
\def\ed{\end{displaymath}}

\def\bc{\begin{center}}
\def\ec{\end{center}}

\def\ba{\begin{array}}
\def\ea{\end{array}}

\begin{document}


\title{The electrostatic instability for realistic pair distributions in blazar/EBL cascades}



\author{S. Vafin$^1$,  I. Rafighi$^{1}$, M. Pohl$^{1,2}$, and J. Niemiec$^{3}$}
\affil{1 Institute for Physics and Astronomy, University of Potsdam, D-14476 Potsdam, Germany\\
2 DESY, Platanenallee 6, D-15738 Zeuthen, Germany\\
3 Instytut Fizyki J\c{a}drowej PAN, ul. Radzikowskiego 152, 31-342 Krak\'{o}w, Poland}

\begin{abstract}
This work revisits the electrostatic instability for blazar-induced pair beams propagating through IGM with the methods of linear analysis and PIC simulations. We study the impact of  the 
realistic distribution function of pairs resulting from interaction of high-energy gamma-rays with the extragalactic background light. We present analytical and numerical calculations of the linear growth rate of the instability for arbitrary orientation of wave vectors. Our results explicitly demonstrate that the finite angular spread of the beam dramatically affects the growth rate of the waves, leading to fastest growth for wave vectors quasi-parallel to the beam direction and a growth rate at oblique directions that is only by a factor of 2-4 smaller compared to the maximum. To study the non-linear beam relaxation, we performed 
PIC simulations that take into account 
a realistic wide-energy distribution of beam particles. The parameters of the simulated beam-plasma system provide an adequate physical picture that can be extrapolated to realistic blazar-induced pairs. In our simulations the beam looses only 1\% percent of its energy, and we analytically estimate that the beam would lose its total energy over about $100$ simulation times. Analytical scaling is then used to extrapolate to the parameters of realistic blazar-induced pair beams. We find that they can dissipate their energy slightly faster by the electrostatic instability than through inverse-Compton scattering. The uncertainties arising from, e.g., details of the primary gamma-ray spectrum are too large to make firm statements for individual blazars, and an analysis based on their specific properties is required.

\end{abstract}




\section{Introduction}\label{Intro}

New Cerenkov telescopes (HESS, VERITAS, MAGIC) and satellites (e.g., Fermi) have discovered many blazars as sources of very high-energy gamma rays ($E\geq 100$ GeV) \citep{Naurois}. Interacting with the extragalactic background light (EBL), these very energetic photons create electron-positron pairs that in turn produce an inverse-Compton cascade in the GeV band. Observations with Fermi-LAT indicate \citep{Neronov10} that the GeV gamma-ray flux from some blazars is lower than that predicted from the full electromagnetic cascade \citep{Neronov09}. One possible explanation is pair deflection in intergalactic magnetic field of strength $B\geq 3\times 10^{-16}$ Gauss \citep{Neronov10,Taylor11}.  In galaxy clusters and in cosmological filaments in general we expect the magnetic field to be stronger than that, and so in filaments the cascades should be reasonably well isotropized and contribute to the extragalactic gamma-ray background. The bulk of the energy transfer to the cascade occurs at distances $D_\gamma\gtrsim 20\ \mathrm{Mpc}$ and hence likely in a cosmological void, where the existence of a fG-level magnetic field is not obvious. An alternative proposal involves beam dissipation by plasma instabilities \citep{Breizman74,Breizman90,Bret04,Bret05,Bret06,Bret10,Godfrey75,Lominadze79}. 

Having established in our earlier paper \citep{Rafighi17} parameter regimes for PIC simulations that permit extrapolation to the very low density pair beams in intergalactic space, we now draw our attention to the impact of realistic beam distribution functions. We present results of both the linear analysis and PIC simulations.


Several authors presented analytical studies of the electrostatic instability of blazar-induced pair beams. \citet{Broderick12} and \citet{RS12} considered a mono-energetic beam with no momentum spread, for which the electrostatic mode reaches its maximum growth rate at a wave vector quasi-perpendicular to the beam direction. This so-called reactive regime is not relevant for realistic blazar-induced beams that have broad energy distributions and a finite angular spreads.

\citet{Miniati13} analyzed the electrostatic mode for a steady-state pair distribution given by the balance between the pair production and inverse Compton losses. They found the maximum growth rate in a direction almost parallel to the beam, in contrast to the case of a cold beam \citep{RS12}. Note that the realistic distribution function is highly non-mono-energetic, and there is no simple relation to the growth rate for a cold beam that depends on its Lorentz-factor ($\sim 1/\Gamma$). Therefore, the growth rate calculated for a mono-energetic beam should not apply to a highly non-mono-energetic plasma. 

\citet{RS13} analyzed the parallel electrostatic mode for a beam with realistic energy distribution of blazar-induced pair-beams \citep{RS12a} in two cases: with and without momentum spread perpendicular to the beam. In contrast to \citet{Miniati13}, they found a negligible effect of the perpendicular momentum spread. We note that \citet{RS13} did not include inverse Compton (IC) losses of the beam in their model and used quite unrealistic energy spectrum of EBL (see below for more details).   

We revisit the pair distribution function resulting from interactions of high-energy gamma rays with EBL photons. The energy distribution of high-energy gamma rays is modeled by a power-law distribution $\propto E^{-1.8}$ \citep{Abdo10}, whereas the EBL energy spectrum is a combination of the model of stellar radiation by \citet{Finke10} and the cosmic X-ray background radiation (CXB) \citep{Fabian92}. Since we investigate, whether or not plasma instabilities can modify the beam faster than the Comptonization would, we ignore inverse-Compton losses which can be relevant only if Comptonization is the fastest process. It will be shown below that the electrostatic instability indeed acts much faster than the IC scattering and its growth rate is much larger than that found by \citet{Miniati13} taking into account the effect of IC scattering.

Using 
the realistic pair distribution, we explicitly demonstrate that the perpendicular momentum spread 
plays a decisive role in shaping the electrostatic mode. 
In particular, we show that without a perpendicular spread the maximum growth rate is observed for wave vectors quasi-perpendicular to the beam direction. However, for a finite perpendicular momentum spread, the maximum growth rate shifts to the direction parallel to the beam and its value is considerably reduced, if compared to that for beams with no perpendicular momentum spread. Moreover, we demonstrate that the maximum growth rate in a slightly oblique direction to the pair beam is much larger than the strictly parallel growth rate found by \citet{RS13}.

At oblique directions of the wave vector the growth rate is only a factor of 3-5 smaller than in the parallel direction, which can be crucial for the instability evolution at the non-linear stage in which non-linear damping may be at play.  We will explore this issue with PIC simulations. Previous work \citep{Rafighi17} describes a range of parameters for which the beam evolution should correctly reflect the physics of the realistic beam and that at the same time can be simulated numerically. Our earlier results, as well as those of \citet{Sironi14} and \citet{Kempf16}, are based on a mono-energetic beam, whereas the realistic pair beam distribution is highly non-mono-energetic. In the current study, we account for a realistic energy distribution to clarify its impact on the non-linear beam evolution.   
 
The paper is organized as follows. Section~\ref{AnModel} presents the stability analysis of the electrostatic mode. Section~\ref{secPIC} describes the results of PIC simulations. Section~\ref{Summary} contains summary and discussion of our results.


\section{Realistic pair spectrum}\label{Pairspectrum}

Before we investigate the growth rate of the electrostatic instability, it is necessary to evaluate the energy distribution of realistic blazar-induced pairs.  Let us consider a fiducial source of high-energy photons with spectrum $F(E_\gamma,z)=dN_\gamma/dE_\gamma$. Of interest are BL Lac objects with intrinsic emission spectrum harder than $E^{-2}$ that extends to at least a few TeV, otherwise the cascade emission would be sub-dominant and would not fall into the energy band accessible with the \textit{Fermi}-LAT. Only a few of the AGN known in the GeV band qualify \citep{Abdo10}. For our fiducial BL Lac we assume a simple power-law spectrum observable at Earth:
\begin{multline}
F(E_\gamma,z=0)= \left(10^{-9}\ \mathrm{\frac{ph.}{cm^2\,s\,GeV}}\right)\, \l( \frac{E_\gamma}{\mathrm{GeV}}\r)^{-1.8}\times \\
\Theta((E_\gamma-E_{\gamma,min})(E_{\gamma,max}-E_\gamma)),
\label{rdf8}
\end{multline}
where $E_{\gamma,min}=0.5$~GeV and $E_{\gamma,max}=50$~TeV. The low-energy limit is irrelevant for the paper, and the high-energy limit is chosen high to explore a case with high pair-beam density and hence strong driving of plasma instabilities. The differential flux at 1 TeV corresponds to 15\% of that of the Crab nebula and is typical of BL Lacs in flaring state \citep{2009ARA&A..47..523H}. The AGN is placed at redshift $z=0.15$, corresponding to a luminosity distance $D_\mathrm{L}\simeq 720$~Mpc.

Gamma-ray emission from jets of AGN has a finite opening angle, and at the site of pair production, at the distance $D_\gamma$ from the AGN, the gamma-ray flux is
\begin{equation}
F(E_\gamma,D_\gamma)= \frac{D_\mathrm{L}^2}{D_\gamma^2}\, F(E_\gamma,z=0)\ ,
\label{rdf8a}
\end{equation}
where the energy is measured at $z=0$. As the density of the pair beam scales with that of the primary gamma rays, it is important to establish at what distance the bulk of pair production happens. Primary gamma rays with $E_\gamma =40$~TeV produce cascade emission above 100~GeV, where the sensitivity of \textit{Fermi}-LAT deteriorates. They also get absorbed within a few Mpc from the AGN, i.e. close to or in a cosmological filament where magnetic field stronger than the fG-level likely exists. Gamma rays with $E_\gamma =10$~TeV have a mean free path of roughly 80~Mpc, i.e. pair-produce in voids, and provide well-detectable cascade emission around 10~GeV. Then, pair beams produced at $D_\gamma\simeq 50$~Mpc represent the best test case for the role of electrostatic instabilities in voids, because one considers the highest beam densities at the location of main energy transfer under the constraint of cascade visibility. For our fiducial AGN, the total number of gamma rays then is
\begin{equation}
N_\gamma= \int dE_\gamma\ F(E_\gamma,D_\gamma=50\,\mathrm{Mpc})\simeq 1.5\cdot 10^{-17}\ .
\end{equation}
Note that at $D_\gamma\simeq 50$~Mpc absorption has already diminished the gamma-ray flux above 10~TeV.

The spectrum of the EBL is denoted as $f(\epsilon,z)=dn_\gamma/d\epsilon$. We define a small redshift interval, $\delta z$, for which $z_b(i)$ ($i=0,1,2...$) represents the boundaries and $z_c(i)$ the center of the interval. In this redshift interval, the evolution of $F(E_\gamma,z)$ can be linearized as \citep{Fang10}
\begin{multline}
F(E_\gamma,z_b(i+1))=F(E_\gamma,z_b(i))e^{-\delta\tau(E_\gamma,z_c(i))} \\
\approx F(E_\gamma,z_b(i))[1 -\delta\tau(E_\gamma,z_c(i))],
\label{rdf1}
\end{multline}
where the optical depth for the small step $\delta z$ in redshift is
\be
\delta\tau(E_\gamma,z)=c\delta z {dt\over dz} \int_0^2 dx {x\over2} \int_{e_th}^\infty d\epsilon f(\epsilon,z) (1+z)^3\sigma_{\gamma\gamma}(\beta).
\label{rdf2}
\ee
Here,
\begin{multline}
{dt\over dz}= {1\over H_0(1+z)}\times \\
\l[ (1+z)^2(1+\Omega_m z) -z(z+2)\Omega_\Lambda\r]^{-1/2},
\label{rdf3}
\end{multline}
with Hubble constant $H_0=70$ km s$^{-1}$ Mpc$^{-1}$, $\Omega_m=0.3$, $\Omega_\Lambda=0.7$, $c$ is the speed of light, while 
\begin{multline}
\sigma_{\gamma\gamma}(\beta)= {3\sigma_T\over 16}(1-\beta^2)\times \\
\l[ 2\beta(\beta^2-2) + (3-\beta^4)\ln\l(1+\beta\over 1-\beta\r)\r]
\label{rdf4}
\end{multline}
describes the cross-section for pair production, where the Thompson cross-section $\sigma_T=6.65\times10^{-25}$ cm$^{2}$, $\beta=(1-4m_e^2c^4/s)^{1/2}$, $s=2E_\gamma\epsilon x(1+z)$, $x=1-\cos\theta$, and $m_e$ is the electron mass. 

The distribution function of newly produced electrons and positrons can be expressed as \citep{Fang10}
\begin{widetext}
\begin{multline}
\delta f_b(\gamma,z_b(i+1))={dn_b\over d\gamma}=c\delta z{dt\over dz}\, \frac{(1+z_c(i))^2}{4}\int_0^\infty dE_\gamma F(E_\gamma,z_b(i)) E_\gamma^2
\int_1^\infty d\gamma^* \times \\ \int_{\gamma^-}^{\gamma^+} 
{d\gamma_{CM}\over \sqrt{\l( {\gamma^*}^2 - 1\r) \l( \gamma_{CM}^2 -1\r)} } 
 \frac{1}{\gamma_{CM}^5}
\delta\l( \gamma^*- {\gamma_{max}\over \gamma_{CM}} \r) 
\int_{\epsilon_{min}}^\infty {d\epsilon\over \epsilon^2} f(\epsilon,z_c(i)) \sigma_{\gamma\gamma}(\beta),
\label{rdf5}
\end{multline}
\end{widetext}
where $\gamma^{\pm}=\gamma\gamma^*\pm \sqrt{\l( {\gamma^*}^2-1 \r) \l(\gamma^2 -1 \r) }$, $\beta=\sqrt{1-\gamma_{CM}^2/\gamma_{max}^2}$, $\gamma_{max}= E_\gamma(1+z_c(i))/(2m_ec^2)$, and $\epsilon_{min}=E_\gamma(1+z_c(i))/(4\gamma_{CM}^2)$. Here, $\gamma_{CM}$ is the gamma-factor of the center-of-mass frame of the interacting photons and $\gamma^*$ is the gamma-factor of pairs in that frame. Integrating Eq. (\ref{rdf5}) over $\gamma_{CM}$ and introducing the 
new variable $x=\gamma_{max}/\gamma^*$, Eq. (\ref{rdf5}) becomes
\begin{widetext}
\begin{multline}
\delta f_b(\gamma,z_b(i+1))= c\delta z{dt\over dz}\, \frac{(1+z_c(i))^2}{4}\,\times \\  
\int_0^\infty dE_\gamma\ F(E_\gamma,z_b(i)) E_\gamma^2  
\int_0^{\gamma_{max}} dx\  \frac{\Theta\l( \l(x- \gamma^-\r) \l(\gamma^+-x\r)\r)}{x^5 \sqrt{\l( (\gamma_{max}/x)^2 - 1\r) \l( x^2 -1\r)}}   \int_{\epsilon_{min}}^\infty {d\epsilon\over \epsilon^2} f(\epsilon,z_c(i)) \sigma_{\gamma\gamma}(\beta),
\label{rdf7}
\end{multline}
\end{widetext}
where $\Theta$ is the Heaviside step function, $\beta=\sqrt{1-x^2/\gamma_{max}^2}$, $\epsilon_{min}=E_\gamma(1+z_c(i))/(4 x^2)$, and $\gamma^\pm= \gamma\gamma_{max}/x \pm \sqrt{(\gamma^2-1)((\gamma_{max}/x)^2-1)}$. Thus, for given $F(E_\gamma,z_b(i))$  and $f(\epsilon,z_c(i))$, Eq. (\ref{rdf7}) yields the increment of the pair distribution function. 

\begin{figure}
\includegraphics[width=86mm,height=70mm]{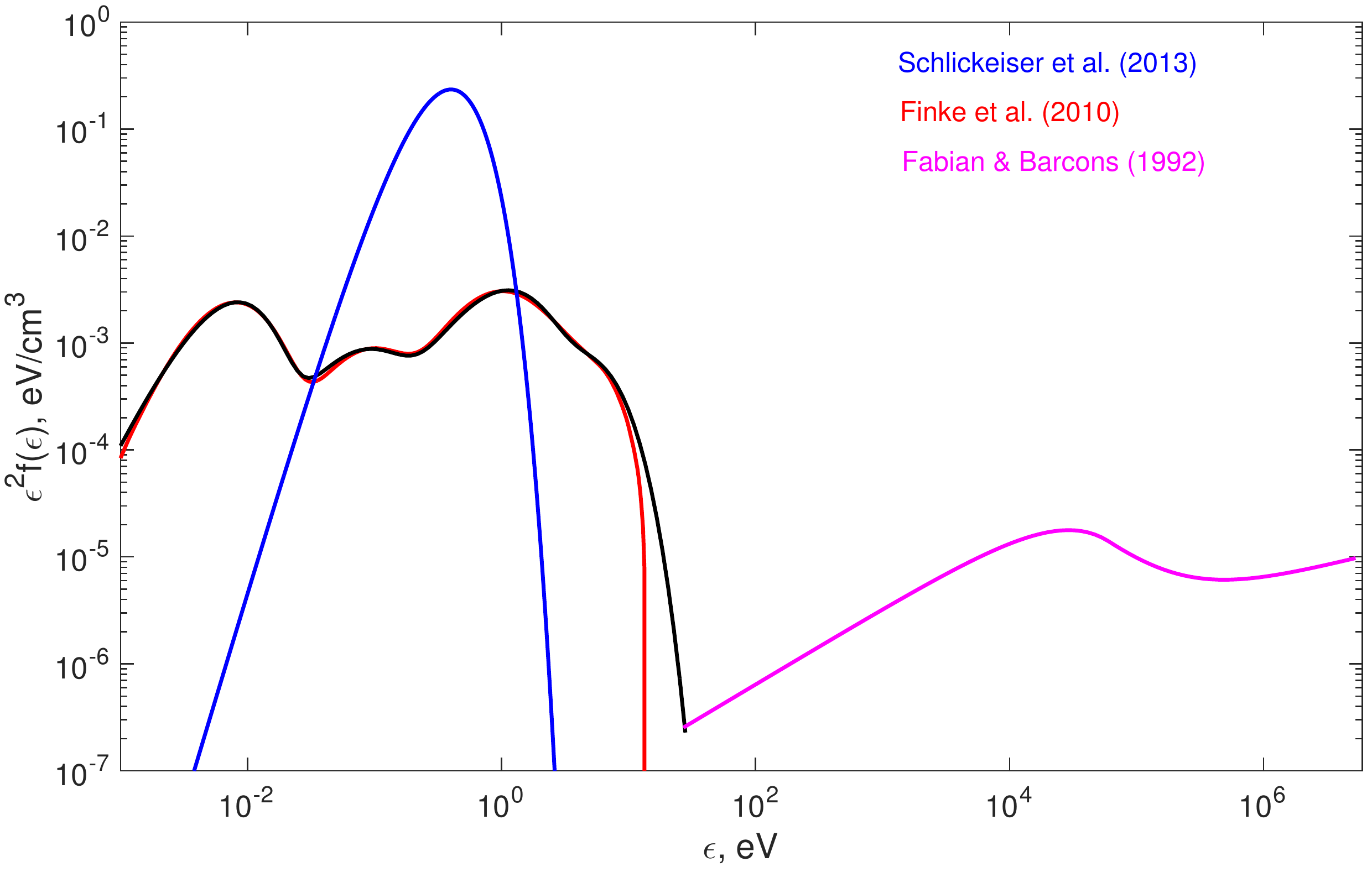}
\caption{EBL energy density spectrum for different models. Blue: The Wien-type distribution used by \citet{RS13} (see also Eq. \ref{rdf9}). Red: Stellar radiation model by \citet{Finke10} at redshift 0.2. Black: Our analytical approximation for the red line (Appendix A). Magenta: empirical model of the X-ray background \citep{Fabian92}.}
\label{EBL_Finke_CXB_spectrum}
\end{figure}

The energy spectrum of low-energy EBL photons is modeled by a combination of stellar radiation and cosmic X-ray background. For the stellar radiation, we used results by \citet{Finke10} for the redshift $0.2$, also available on-line\footnote{http://www.phy.ohiou.edu/$\sim$finke/EBL/index.html}, whereas the X-ray background radiation is described by empirical fits found by \citet{Fabian92}. Note that \citet{RS13} used an EBL spectrum of Wien-type,
\be 
f(\epsilon)= {N_0\over \Gamma(3) k_B T_W} \l( \epsilon \over k_B T_W\r)^2 \exp\l({- {\epsilon\over k_B T_W}}\r),
\label{rdf9}
\ee
where $\Gamma$ denotes the Gamma function, $k_BT_W=0.1$ eV, and $N_0=1$ cm$^{-3}$.
Fig. \ref{EBL_Finke_CXB_spectrum} compares model spectra of the EBL. It is obvious that the Wien-type distribution Eq. (\ref{rdf9}) is not a good description of the realistic EBL spectrum given by stellar radiation and CXB. Thus, in the present work, we will use an analytical approximation (see Appendix A) for the spectral models by \citet{Finke10} (shown as red curve in Fig. \ref{EBL_Finke_CXB_spectrum}) along with the approximations by \citet{Fabian92}.
Electron-positron pairs produced by gamma-ray absorption will eventually loose their energy by Comptonization of the microwave background, unless plasma instabilities drain their energy more rapidly. In any case, the mean free path for Comptonization,
\begin{equation}
\lambda_\mathrm{IC}\simeq (75\ \mathrm{kpc}) \left(\frac{10^7}{\gamma}\right)
\end{equation}
provides the upper limit for the path length along which the pair beam can build up. A 10-TeV gamma ray produces electrons and positrons with mean energy 5~TeV, or $\gamma=10^7$, and re-radiates gamma rays with energies around 10 GeV that should be easily observable with the \textit{Fermi}-LAT. We use the pathlength $\lambda=\lambda_\mathrm{IC} (\gamma=10^7)$ to calculate the accumulated pair spectrum.

Fig. \ref{Pairspectrum_over_nb_1,8Mpc} shows the pair spectrum at distance 50~Mpc from a blazar resulting from interactions of high-energy gamma rays with a spectrum following Eq. (\ref{rdf8}) and low-energy EBL photons \citep{Finke10,Fabian92}. The total number density of pairs is about $3\cdot10^{-22}$ cm$^{-3}$. The red and the blue curve illustrate results of our calculation (Eq. \ref{rdf9}), while the black one shows the approximation found by \citet{RS12a},
that is close to our results only beyond the peak at $\gamma>10^6$. The rising flank of the black curve at $\gamma\approx 10^{4.7}$ is much steeper and it does not show a second peak at lower energies. It is clear that the peak of the pair distribution at $\gamma\approx 5\times 10^2$ results from the cosmic X-ray background radiation which was not included in the model by \citet{RS13}. In Appendix B, we give an analytical approximation for the red curve in Fig. \ref{Pairspectrum_over_nb_1,8Mpc} that we use in the next sections to evaluate the linear growth rate of the electrostatic instability.  
\begin{figure}
\includegraphics[width=86mm,height=70mm]{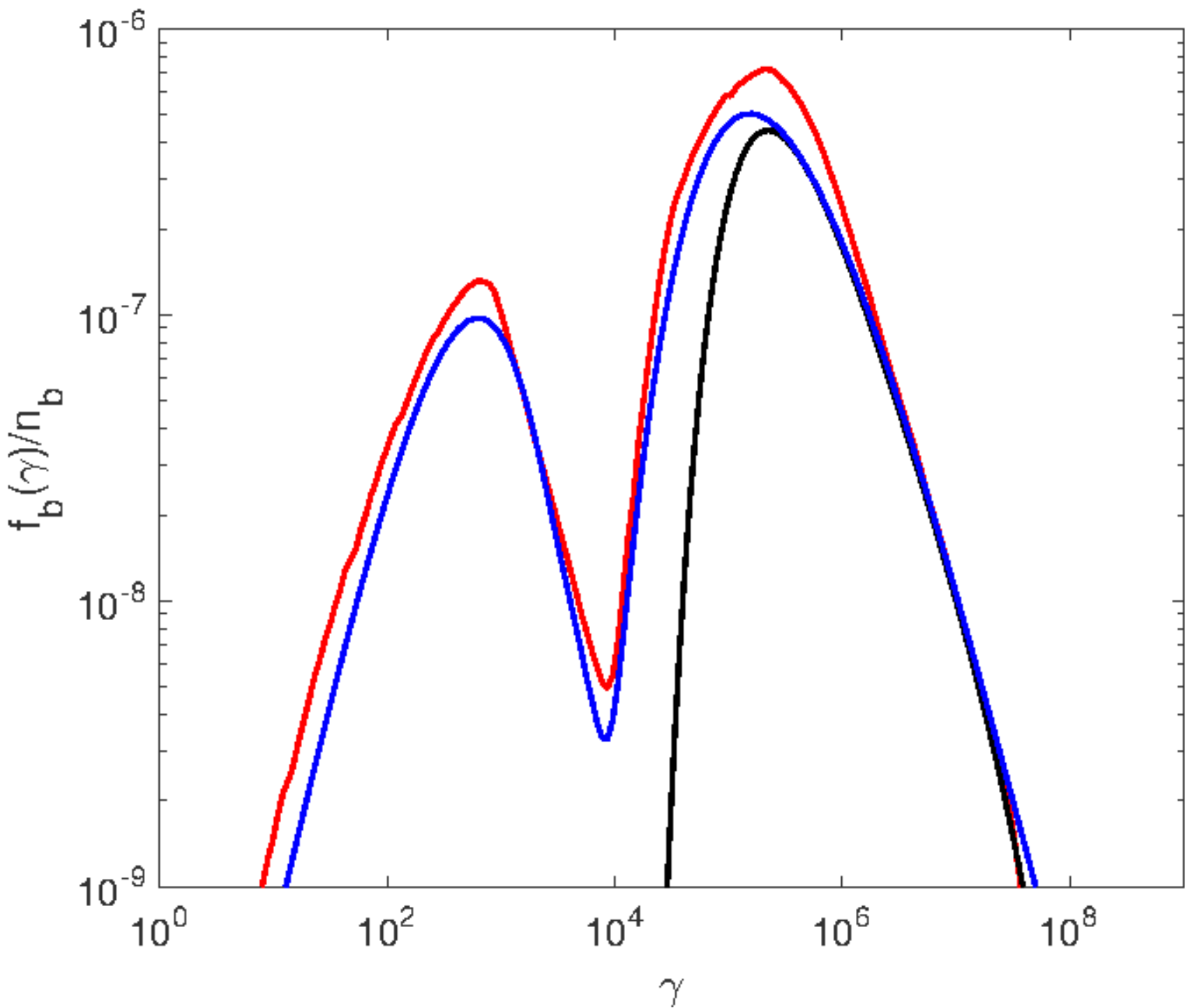}
\caption{Normalized pair spectrum at the distance 50 Mpc from the source. Red: Numerical solution of Eq. (\ref{rdf7}) using the Finke model model and the CXB fit. Blue: our approximation for the red curve (Eq.~(\ref{app2})). Black: Pair spectrum for the EBL approximation by \citet{RS13}.}
\label{Pairspectrum_over_nb_1,8Mpc}
\end{figure}

\section{Growth rate of electrostatic instability}\label{AnModel}

\subsection{Linear growth rate}

Since the growth rate of the electrostatic instability is mainly regulated by the momentum distribution of the beam particles, we introduce the normalized momentum distribution function, $f_b({\bf p})=f_b({\bf p, \bf x})/n_b$. The background plasma is assumed to be cold. Then the dispersion equation for the electrostatic mode reads \citep{Breizman90}

\be
\Lambda({\bf k},\omega)=1-{\omega_p^2\over\omega^2}- \sum_b {4\pi n_b e_b^2\over k^2} \int d^3p { {\bf k} {\pa f_b({\bf p})\over\pa{\bf p}} \over {\bf k}{\bf v}-\omega }=0,
\label{am1}
\ee
where $n_b$ is the density, and $e_b$ the charge, of the beam particles, $\omega_p=(4\pi n_p e^2/m_e)^{1/2}$ is the plasma frequency of the IGM of density $n_p$, and ${\bf k}=(k_\perp,0,k_z)$ denotes the wave vector. In our case, the beam is composed of electrons and positrons ($e_b=e$). The ratio $n_b/n_p\ll 1$ is small, and we can find the growth rate $\omega_i=\Im\omega=\Im(\omega_r+i\omega_i)$ as
\be
\omega_i= - {\Im\Lambda(\omega=\omega_r) \over {\pa\Re\Lambda(\omega=\omega_r)\over\pa\omega_r}},
\label{am2}
\ee
where
\be
\Re\Lambda(\omega=\omega_r)\approx 1- {\omega_p^2\over\omega_r^2},
\label{am3}
\ee
and
\begin{multline}
\Im\Lambda(\omega=\omega_r)\approx 
- \sum_b {4\pi^2 n_b e_b^2\over k^2} \\
\times \int  {\bf k} {\pa f_b({\bf p})\over\pa{\bf p}}  \delta\l( {\bf k}{\bf v}-\omega_r \r) d^3p.
\label{am4}
\end{multline}
Eqs. (\ref{am2})-(\ref{am3}) yield $\omega_r=\omega_p$ and
\be
\omega_i=\omega_p \sum_b {2\pi^2 n_b e_b^2\over k^2} \int  {\bf k} {\pa f_b({\bf p})\over\pa{\bf p}}  \delta\l( {\bf k}{\bf v}-\omega_r \r) d^3p. 
\label{am5}
\ee
For an ultra-relativistic beam, Eq. (\ref{am5}) yields \citep{Breizman90}
\begin{multline}
\omega_i= \pi {n_b\over n}\omega_p \l(\omega_p\over kc \r)^3\int_{\theta_1}^{\theta_2} d\theta' \times \\
{-2g(\theta')\,\sin\theta' + (\cos\theta' -kc\cos\theta/\omega_p)(dg/d\theta') \over [(\cos\theta'-\cos\theta_1)(\cos\theta_2-\cos\theta')]^{1/2}},
\label{am5.1}
\end{multline}
where 
\be
g(\theta')=m_e c\int_0^\infty pf_b(p,\theta') dp,
\label{am7}
\ee
$\theta$ is the angle between the beam propagation and the wave vector, $\cos\theta_{1,2}=\omega_p[\cos\theta\pm\sin\theta\sqrt{(kc/\omega_p)^2-1}]/(kc)$, and the integration angle $\theta'$ is counted from the beam direction.



\subsection{Electrostatic instability for a beam without perpendicular momentum spread}\label{AnalytRS}



In this section, we consider the linear electrostatic growth rate for a realistic blazar-induced pair beam, but with no transverse angular spread. It was found that the momentum distribution of produced pairs is strongly collimated along the direction of the initial gamma-rays, and the transverse momentum is around $m_ec/2$, leading to an opening angle of the beam of about $10^{-5}-10^{-4}$ \citep{Miniati13}. Given the small transverse component of the momentum it may seem to be reasonable to model the distribution function of a beam propagating along the z-axis as
\be
f_b({\bf p})=f_{b,z}(p_z)\delta(p_x)\delta(p_y),
\label{r1}
\ee
where $f_{b,z}(p_z)$ is related to the distribution found in the previous section (Eq. \ref{app2}) as $f_{b,z}(p_z)=f_b(\gamma\approx p_z/(m_ec))/(m_ec)$. 
Inserting Eq. (\ref{r1}) into Eq. (\ref{am5}), we obtain
\begin{multline}
\omega_i= \omega_p \sum_b {2\pi^2 n_b e_b^2\over k^2} \int_{-\infty}^{\infty} \l[ k_z {\pa f_{b,z}\over\pa p_z}\delta\l( {k_zp_z\over m_e\gamma} -\omega_p\r) - \r. \\ \l. 
-{k_\perp^2 f_{b,z}\over m_e\gamma} \delta'\l( {k_zp_z\over m_e\gamma} -\omega_p\r) \r]dp_z,
\label{r3} 
\end{multline}
where $\gamma=\sqrt{(p_z/m_ec)^2+1}$. Eq. (\ref{r3}) can be rewritten as
\begin{multline}
\omega_i=\pi {n_b\over n_p} \omega_p \gamma_0^3\l(\omega_p\over kc \r)^2 
\l[    2p_{z,0}f_{b,z}(p_{z,0}) \l(k_\perp\over k_z \r)^2 + \r. \\ \l. + (m_ec)^2 \l(1+ \l( \gamma_0 k_\perp\over k_z\r)^2 \r) {\pa f_{b,z}(p_{z,0})\over \pa p_{z,0}} \r],
\label{r4} 
\end{multline}
where $p_{z_0}=m_ec/\sqrt{(k_zc/\omega_p)^2-1}$ and $\gamma_0=\sqrt{(p_{z,0}/m_ec)^2+1}$. For the limiting case of the parallel waves, the growth rate reads
\be
\omega_{i,||}=\pi {n_b\over n_p} \omega_p \gamma_0^3\l(\omega_p\over kc \r)^2 (m_ec)^2 {\pa f_{b,z}(p_{z,0})\over \pa p_{z,0}}.
\label{r5} 
\ee 
However, the growth rate (\ref{r4}) is larger by many orders of magnitude in quasi-perpendicular direction $k_\perp\gg k_z$
\be
\omega_{i,\perp}\approx \pi {n_b\over n_p} \omega_p \gamma_0^5\l(\omega_p\over kc \r)^2 (m_ec)^2 {\pa f_{b,z}(p_{z,0})\over \pa p_{z,0}}=\gamma_0^2\omega_{i,||}.
\label{r6}
\ee

\begin{figure}
\includegraphics[width=86mm,height=70mm]{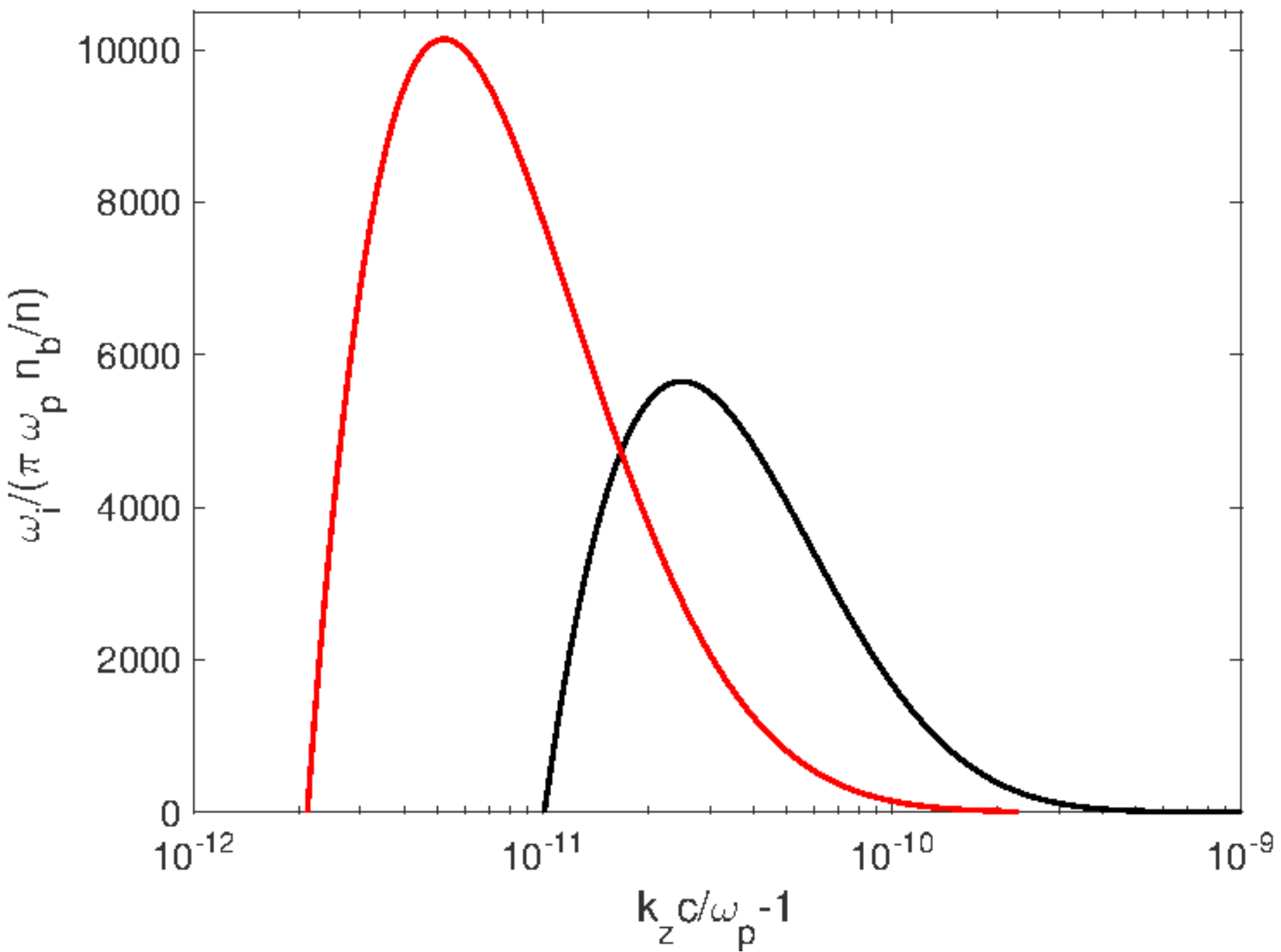}
\caption{Growth rate of the electrostatic instability for parallel wave vectors for beam with no angular spread and distribution function as determined by us (Eq.~\ref{app2}, red line) and \citet{RS12a} (Eq.~\ref{rdf10}, black line), respectively.}
\label{No_angular_spread_ParallelGR_comparison}
\end{figure}

The growth rate for parallel wave vectors (\ref{r5}) is shown in Fig. \ref{No_angular_spread_ParallelGR_comparison} for two different pair spectra, the approximation by \citet{RS12a} (Eq. \ref{rdf10}) and our result (Eq. (\ref{app2})). To be noted from the figure is that the maximum growth rates for parallel wave vectors differ by only a factor 2. 

The growth rate for arbitrary wave vectors is presented in Fig. \ref{No_anglular_spread_RSapp_SpecInd=1.8} and Fig. \ref{No_anglular_spread_EBL_Finke_CXB}, respectively, for pair beams with distributions (\ref{rdf10}) and (\ref{app2}). It is evident that for a more realistic pair distribution (\ref{app2}) the peak of the growth rate is narrower and has a much larger value than for pair distribution (\ref{rdf10}). Since the perpendicular growth rate is proportional to $\gamma_0^2$, the high-energy part of the distribution function (see Fig. \ref{Pairspectrum_over_nb_1,8Mpc}) gives the dominant contribution to the growth rate.    
\mpo{
\begin{figure}
\centering
\subfloat[ ]{
   \includegraphics[width=0.99\columnwidth]{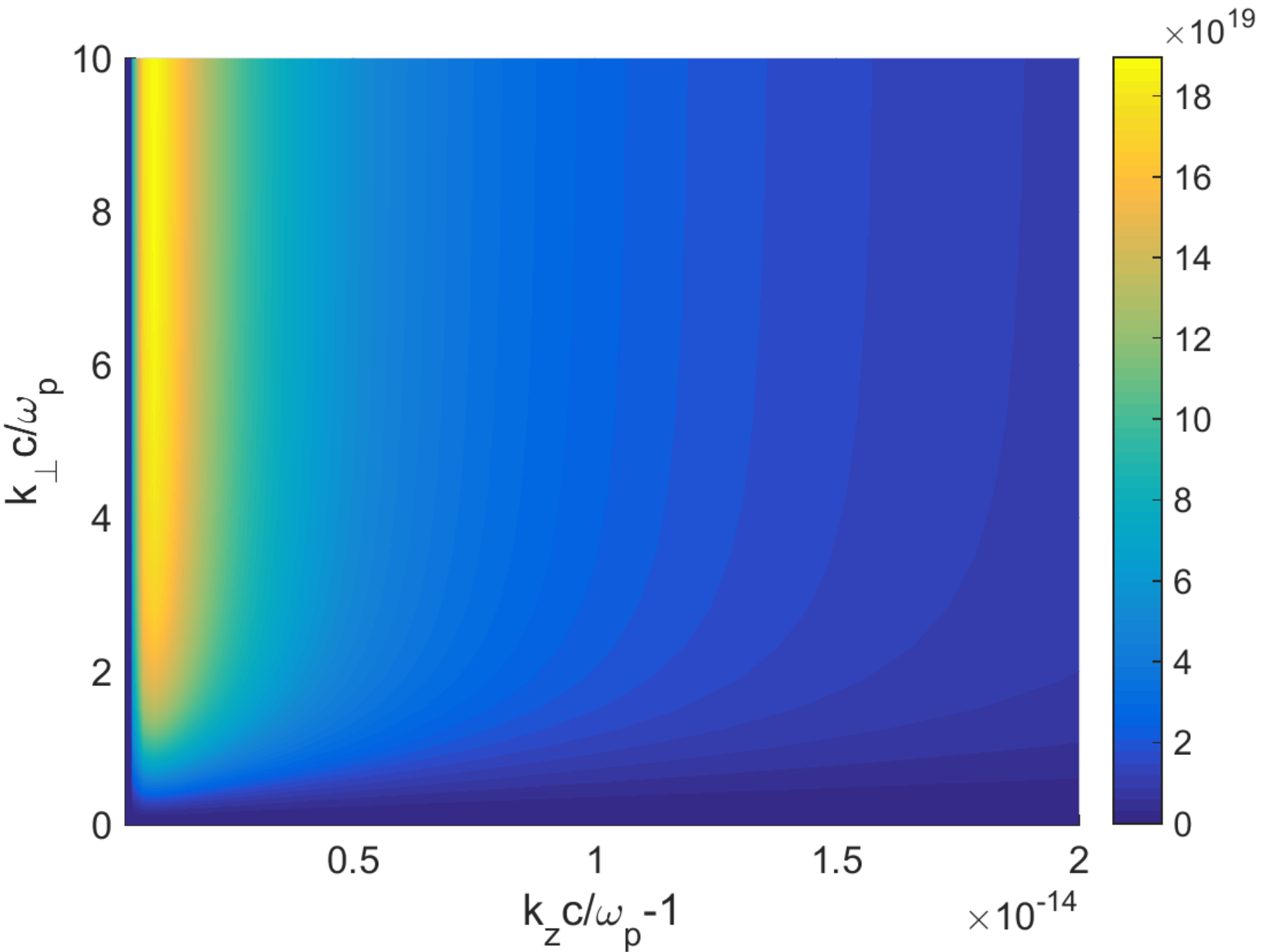}
\label{No_anglular_spread_RSapp_SpecInd=1.8}
}\quad
\subfloat[ ]{
   \includegraphics[width=0.99\columnwidth]{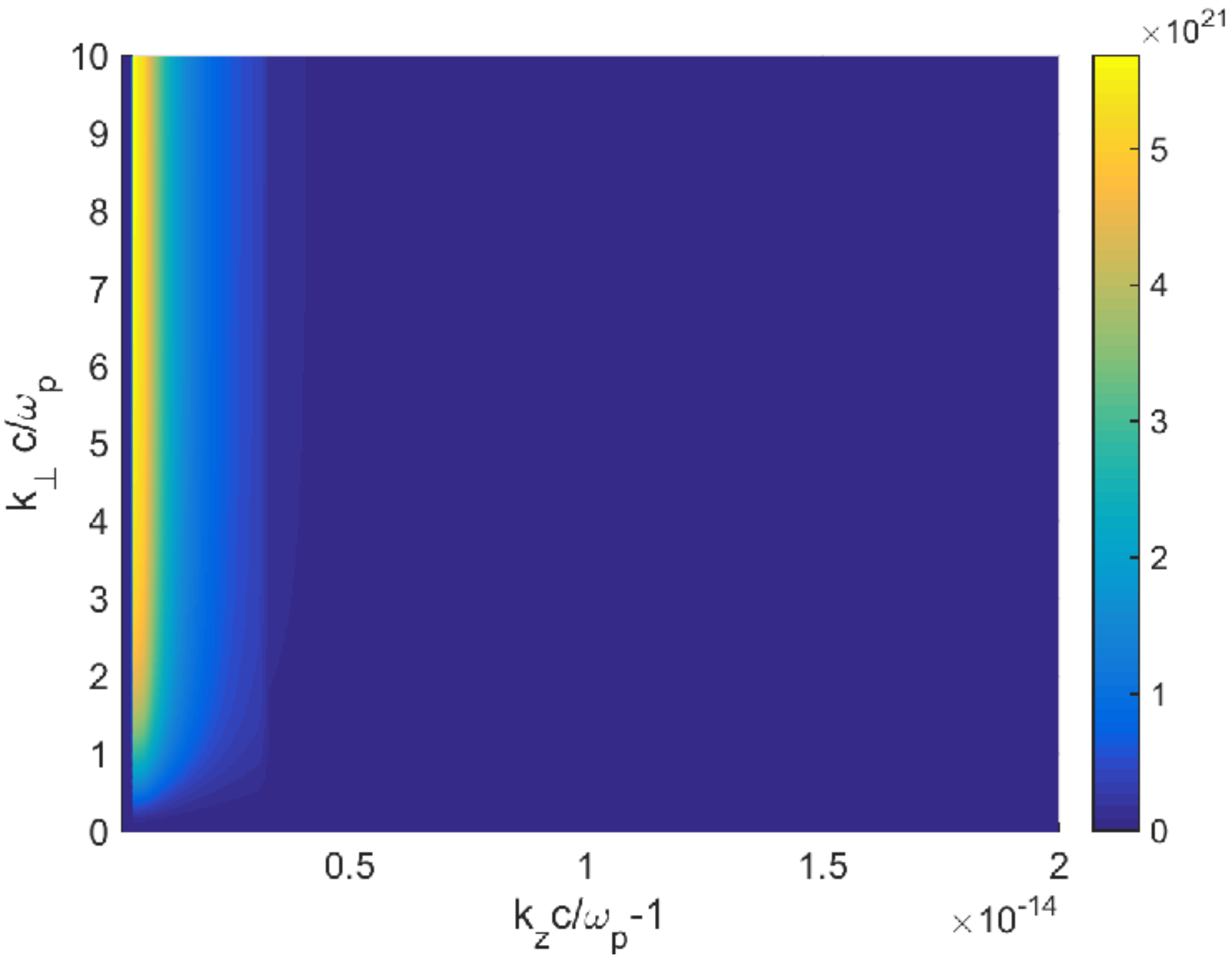}
   \label{No_anglular_spread_EBL_Finke_CXB}
}
\caption{(a) Normalized growth rate, $\omega_i/(\pi \omega_p(n_b/n_p))$, for a beam with no angular spread and distribution function Eq. (\ref{rdf10}) (\citet{RS12a}, $M_c=2\times10^6$, $\tau_0=10^3$, $\alpha=1.8$). (b) As in (a), but for a beam with no angular spread and the distribution function derived by us (Eq.~\ref{app2}).}
\end{figure}
 }

\subsection{Electrostatic instability for a beam with a finite perpendicular momentum spread}\label{AnalytM}
Now, we will include into consideration a finite opening angle of the beam. To do this, we use the distribution function in the form 
\be
f_{b}(p,\theta)= f_{b,p}(p)f_{b,\theta}(\theta,p),
\label{m1}
\ee
where the transverse distribution can be well approximated by \citep{Miniati13}
\be
f_{b,\theta}(\theta,p)\approx \frac{1}{\pi\,\Delta\theta^2} \, \exp \l(-\frac{\theta^2}{\Delta\theta^2 }\r)
\label{m3}
\ee 
and the angular spread for pairs with momentum $p$ can be estimated as $\Delta\theta\approx m_ec/p$ \citep{Broderick12}. The distribution $f_{b,p}(p)$ is derived by transformation from the $z$-integral over Eqs.~(\ref{rdf1}) and (\ref{rdf7}),  
\be 
f_{b,p}(p)=f_b\l(\gamma\approx \frac{p}{m_e c}\r)\,\frac{d\gamma}{p^2dp}=\frac{f_b\l(\gamma\approx \frac{p}{m_e c }\r)}{m_e cp^2}\ .
\ee

\begin{figure}
\centering
\subfloat[ ]{
   \includegraphics[width=0.99\columnwidth]{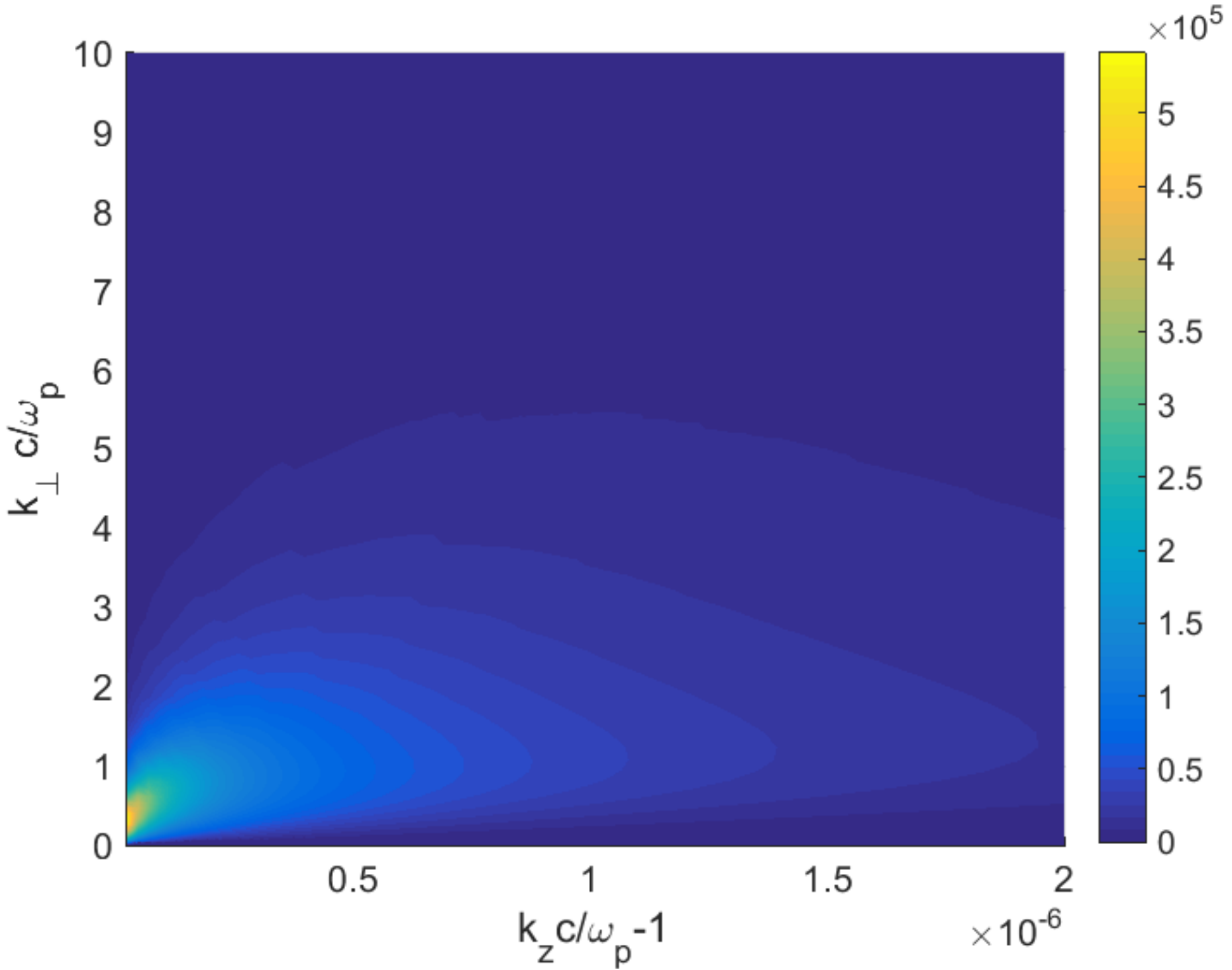}
   \label{Spread_1overGamma_ksi2_calc1_RS} 
}\quad
\subfloat[ ]{
   \includegraphics[width=0.99\columnwidth]{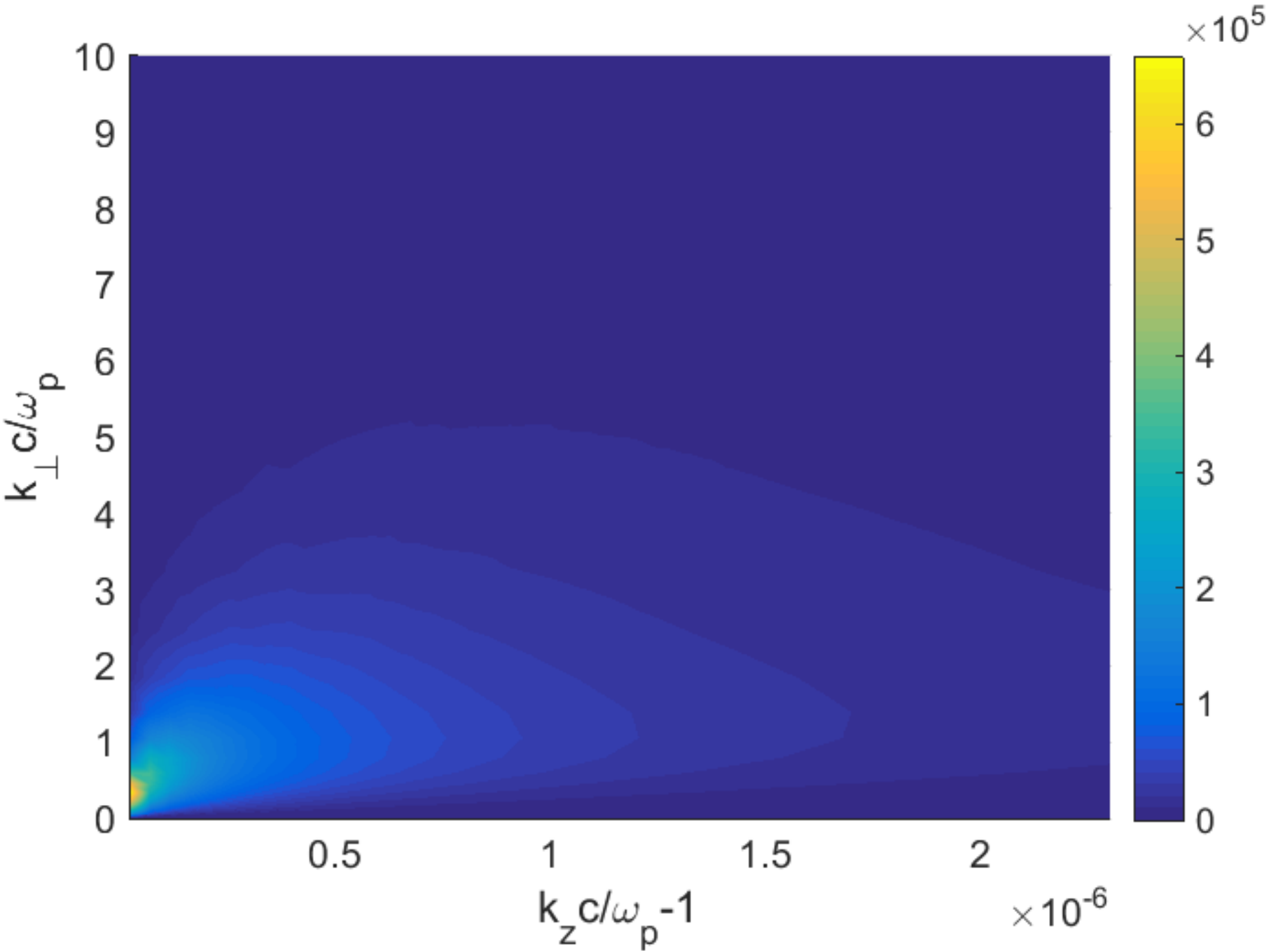}
   \label{Spread_1overGamma_ksi2_calc2}
}
\caption{(a) Normalized growth rate, $\omega_i/(\pi \omega_p(n_b/n_p))$, for a beam with a finite angular spread and distribution function Eq. (\ref{rdf10}). (b) As for (a), but for the distribution function Eq. (\ref{app2}) determined by us.}
\end{figure}

The growth rate of electrostatic waves is evaluated by numerically solving Eq. (\ref{am5}). Fig. \ref{Spread_1overGamma_ksi2_calc1_RS} illustrates the growth rate for a beam with the distribution function given by Eq. (\ref{rdf10}) (Schlickeiser's approximation), whereas Fig. \ref{Spread_1overGamma_ksi2_calc2} is based on the distribution function found by us and approximated by Eq. (\ref{app2}). The distribution of the growth rate in the wave-vector space is very similar for the two pair distributions, and we can conclude that low-energy pairs that are not included in Schlickeiser's approximation do not really matter. 
We find that (i) the growth rate for the beam with a finite angular spread is much smaller compared to the beam with no angular spread and (ii) the maximum growth rate becomes very narrow and lies in the quasi-parallel direction ($\theta\approx16.7^{\circ}$) to the beam. To be also noted is that the growth rate at oblique wave vectors with $\theta\approx 40^{\circ}$ is only by a factor 3-5 smaller than the peak value.

The maximum growth rate is 
\be 
\omega_i^{max}\approx 3.2\times10^{-6} n_{b,20} n_{e,7}^{-0.5} \ \mathrm{s}^{-1}\ ,
\ee 
where we adopted $n_b= n_{b,20}10^{-20}$ cm$^{-3}$ and $n_e=n_{e,7}10^{-7}$ cm$^{-3}$. This growth rate must be compared to the IC loss time, $\tau_{IC}\approx 10^{20}\gamma^{-1}$ s. Thus, even pairs with Lorentz factor $\gamma\approx 10^9$ will drive the instability for thousands of exponential growth cycles before they loose energy to IC scattering. The important questions are at what level the mode saturates and at which rate the beam energy is depleted while the wave mode is at saturation. While we address the former with PIC simulations, the latter requires an analytical estimate.

\section{PIC simulations}\label{secPIC}

%

\subsection{Model description and results}

A realistic blazar-induced pair beam propagating through IGM cannot be simulated numerically due to its very small number density. Nevertheless, a range of beam and plasma parameters can be found such that the problem is numerically accessible with a PIC code, and the physical picture can be extrapolated to the realistic situation. In fact, several conditions must be satisfied \citep{Kempf16,Rafighi17}: 

(i) the beam/plasma energy density ratio must be less than unity;

(ii) the Weibel mode has to be stable; 

(iii) the electrostatic instability should develop in the kinetic regime. 

\begin{figure}[]
\includegraphics[width=84mm]{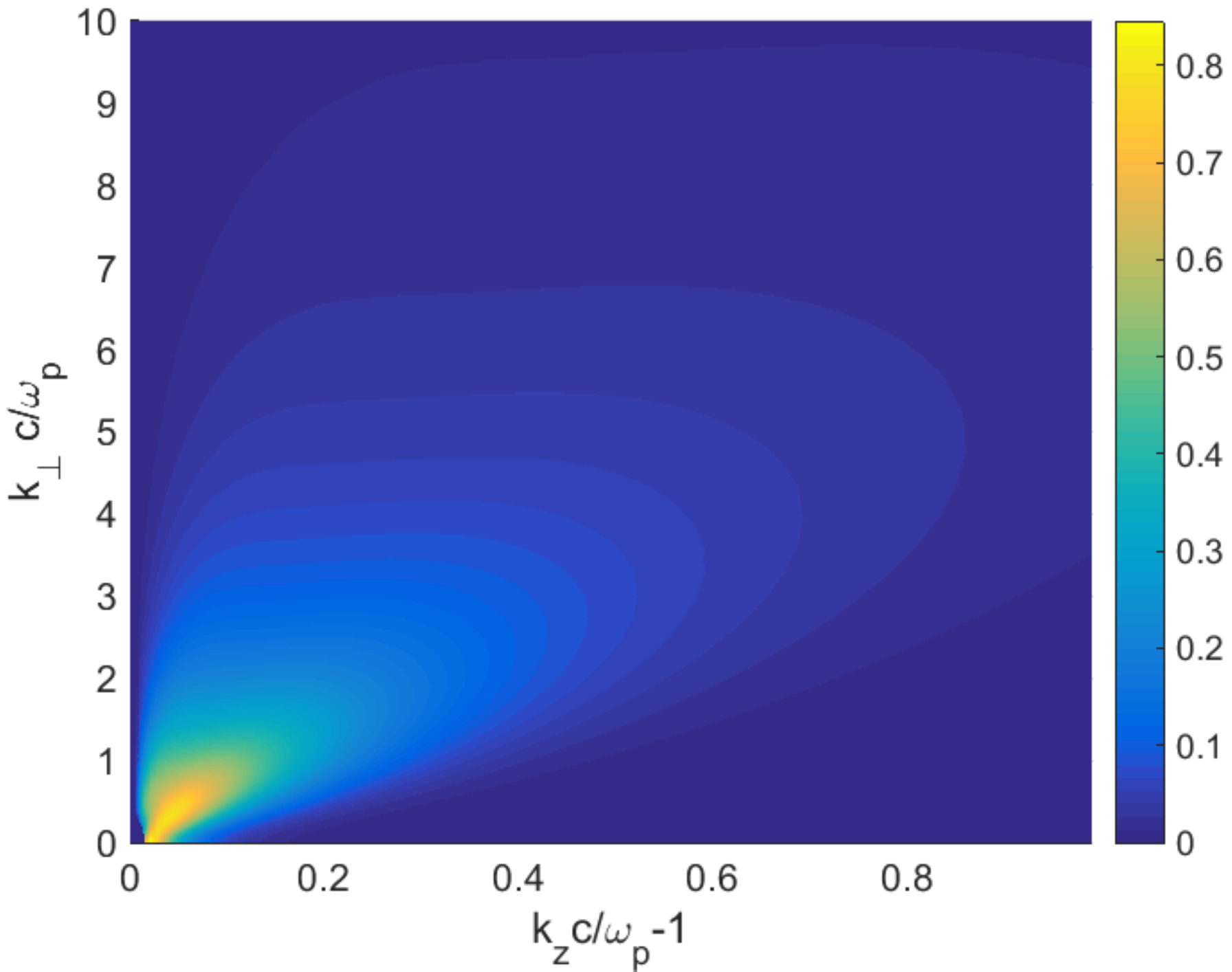}
\caption{Normalized growth rate, $\omega_i/(\pi \omega_p(n_b/n_p))$, for the composite distribution function used in PIC simulations and displayed in Fig. \ref{Distribution_muR=2,5_g1=5(0,9)_g2=20(0,1)}.}
\label{muR_2_5_g1_5_0_9__g2_20_0_1__Calc_no_sing_calc2}
\end{figure}

Earlier simulation studies considered only a mono-energetic Maxwellian beam \citep{Sironi14,Kempf16,Rafighi17} which is not a good representation of the real situation, because the true pair distribution is highly non-mono-energetic. Maxwellian beams are easy to generate in a simulation \citep{Zenitani15}, and an efficient method of inserting a non-Maxwellian beam would be the superposition of two or more Maxwellian beams. As the production of pairs with high center-of-momentum energy is Klein-Nishina-suppressed, and $p_\perp$ is Lorentz-invariant, the Maxwellian beams to be superposed should have the same rest-frame temperature ($kT_b\approx 200$ keV), but will differ in their gamma-factors. Here we present simulations for a composite beam with normalized distribution function
\begin{widetext}
\be
f(p_z,p_\perp)= {w_1\mu_R \over 4\pi (m_ec)^3\Gamma_1 K_2(\mu_R)}e^{-\mu_R\Gamma_1 \l[ \l( 1+{p_z^2+p_\perp^2\over(m_ec)^2} \r)^{1/2}- \beta_1 {p_z\over m_ec}  \r]} + \\
{w_2\mu_R \over 4\pi (m_ec)^3\Gamma_2 K_2(\mu_R)}e^{-\mu_R\Gamma_2 \l[ \l( 1+{p_z^2+p_\perp^2\over(m_ec)^2} \r)^{1/2}- \beta_2 {p_z\over m_ec}  \r]}\ . 
\label{pic1}
\ee
\end{widetext}
We choose for the beam Lorentz factors $\Gamma_1=5$ for beam 1 and $\Gamma_2=20$ for beam 2. Furthermore, we use $\mu_R = m_ec^2/(k_BT_b)$ and $\beta_{1,2}= \l(1-1/\Gamma_{1,2}^2 \r)^{1/2}$. The relative weight factors of the beams are $w_1=0.9$ (beam 1) and $w_2=0.1$ (beam 2). The beam momentum distribution integrated over the transverse momentum is shown in Fig. \ref{Distribution_muR=2,5_g1=5(0,9)_g2=20(0,1)} and designed to resemble the high-energy part of the expected pair distribution displayed in Fig.~\ref{Pairspectrum_over_nb_1,8Mpc}. Our discussion of the linear growth rate in Section~\ref{AnalytM} indicated that it is this high-energy part that matters. We shift it to low Lorentz factors to render the PIC simulations numerically stable. The simulation time is long enough to follow the electrostatic instability, while keeping it  in the kinetic regime. The linear growth rate of the electrostatic instability is displayed in Fig. \ref{muR_2_5_g1_5_0_9__g2_20_0_1__Calc_no_sing_calc2}. The growth rate has its peak value in the quasi-parallel direction, $20-25^{\circ}$ to the beam which approximately reproduces the growth rate for the realistic pair beam (Fig. \ref{Spread_1overGamma_ksi2_calc2}). 

\begin{figure}[]
\includegraphics[width=84mm]{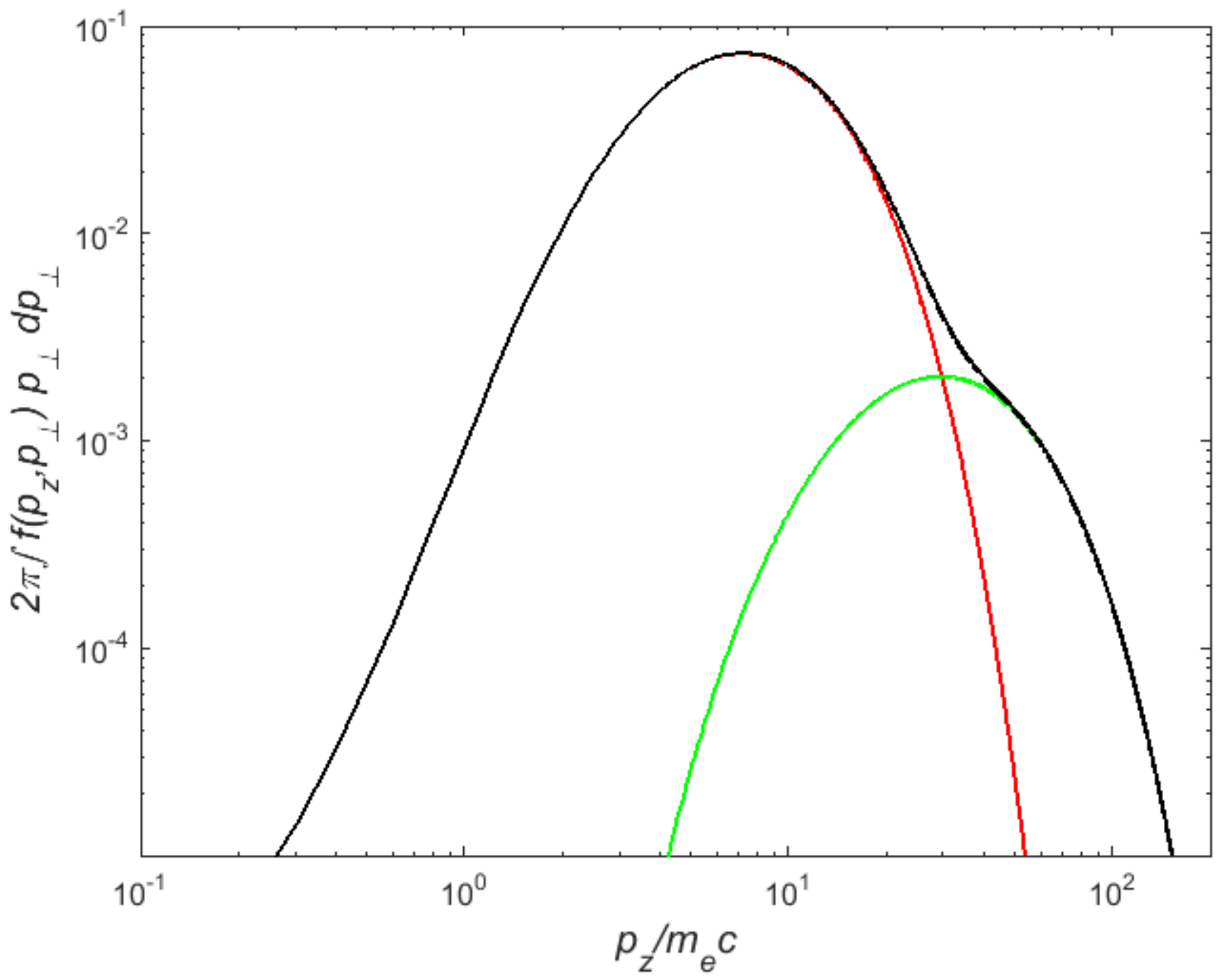}
\caption{Composite beam distribution function used for PIC simulations. Red: beam 1 ($T_b=200$ keV, $\Gamma_1=5$, $w_1=0.9$). Green: beam 2 ($T_b=200$ keV, $\Gamma_2=20$, $w_2=0.1$). Black: the total distribution.}
\label{Distribution_muR=2,5_g1=5(0,9)_g2=20(0,1)}
\end{figure}

We chose the beam/plasma density ratio equal to $\alpha=2\times10^{-4}$ and the background plasma temperature 2 keV. Then the beam/plasma energy density ratio is about $\epsilon=0.66$ which is smaller than 1. Moreover, the Weibel mode is stable, since the condition $\l\langle p_\perp \r\rangle > \l\langle p_\parallel \r\rangle \l(\alpha/\l\langle \Gamma \r\rangle \r)^{1/2}$ is fulfilled for the beam \citep{Rafighi17}. 

For comparison between one- and two-beam systems, we also ran a simulation with $w_1=1$ and $w_2=0$. In this case, $\epsilon=0.5$, the electrostatic instability develops likewise in the kinetic regime, and the Weibel mode is stable \citep{Rafighi17}. The parameters of our simulations are summarized in Table \ref{Tab3}. 

For our PIC simulations, we used the multi-dimensional fully electromagnetic relativistic code EPOCH 2D created by the Collaborative Computational Plasma Physics (CCPP) consortium and supported
by the Engineering and Physical Sciences Research Council (EPSRC) \citep{Arber15}. Additionally, we have introduced the algorithm of \citet{Zenitani15} to generate the relativistic Maxwellian distribution for the beams. The simulation plane was chosen to be z-y plane with periodic boundary conditions. The pair beam propagates along the z-axis through the electron-proton plasma. The beam and plasma particles have the real mass ratio and fill the whole simulation box. The density ratio, $\alpha=2\times10^{-4}$, is set with numerical weights. The simulation box is presented by $2048 \times512$ cells, each 1/4 of the skin length in size,  $ \lambda_{e}=\frac{c}{\omega_{pe}}=4\,\Delta_z$. Tests demonstrated a sufficient suppression of self heating and statistical noise for our simulation setup that involves 400 computational particles per species and cell, a 6th-order field particle pusher, and a triangular-shaped cloud (TSC) shape function. The size of the simulation box and the choice of skin length also provide sufficient resolution for grid frequencies of the narrow resonance at which the electrostatic instability operates \citep{2017arXiv170400014S}.   
 
\begin{table}[]
\caption{Simulation parameters for composite beams with $\Gamma_1=5$ and $\Gamma_2=20$.} 
\centering 
\begin{tabular}{|lc|c|c|} 
\hline 
Parameter &  & run 1 & run 2 \\ [0.5ex]
\hline  
Density ratio & $\alpha$ & $2\times10^{-4}$ & $2\times10^{-4}$  \\
\hline
Plasma temperature & $T_p$ & 2 keV & 2 keV \\
\hline
Beam temperature &$T_b$ & 200 keV & 200 keV  \\
\hline
Weight for beam 1 & $w_1$ & 1 & 0.9  \\
\hline
Weight for beam 2 & $w_2$ & 0 & 0.1  \\
\hline
Energy density ratio & $\epsilon$ & 0.5 & 0.66 \\ [1ex]
\hline
\end{tabular}
\label{Tab3} 
\end{table}

\begin{figure}[]
\includegraphics[width=84mm]{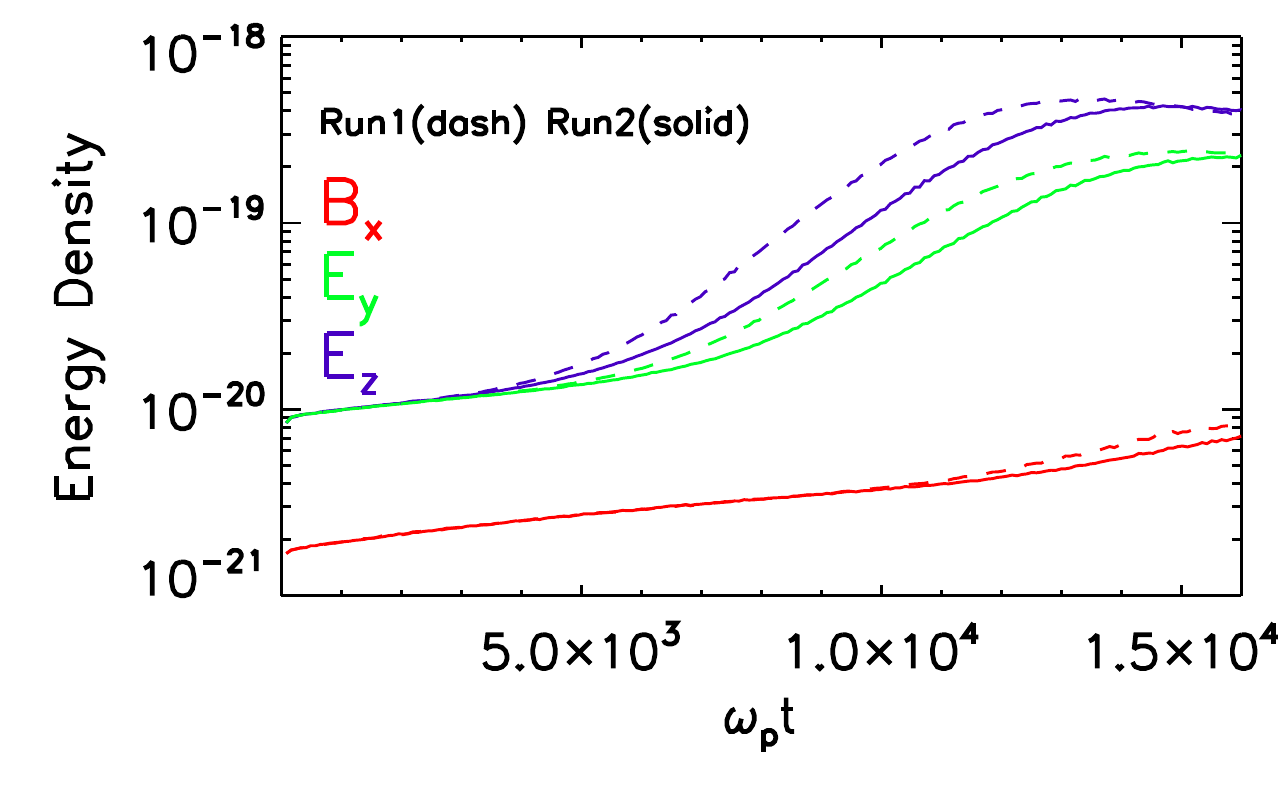}
\caption{Time evolution of the energy densities of electric and magnetic field, respectively, in SI units for run 1 (dashed lines) and run 2 (solid lines). }
\label{energt3g5}
\end{figure}

Fig. \ref{energt3g5} compares the energy-density evolution of the electric and magnetic fields for runs 1 and 2. The linear growth rate of the electric field in the simulations agrees quite well with the theoretical calculations providing $\omega_i=10^{-3}\omega_p$ for run 1 and $\omega_i=9\times10^{-4}\omega_p$ for run 2, at least during the initial phase. Already after $5000 \omega_p^{-1}$ wave growth proceeds at a reduced pace. The numerical experiments also reproduce the slightly faster growth in run 1 compared to run 2. Although the second, high-energy beam provides additional high-energy particles, their effect on the growth rate is negligible. Thus, the growth rate is mainly determined by beam 1, and the slightly higher density ratio, $\alpha$, in run 1 leads to a marginally larger growth rate of the electrostatic instability.

\begin{figure}
\centering
\subfloat[ ]{
   \includegraphics[width=0.99\columnwidth]{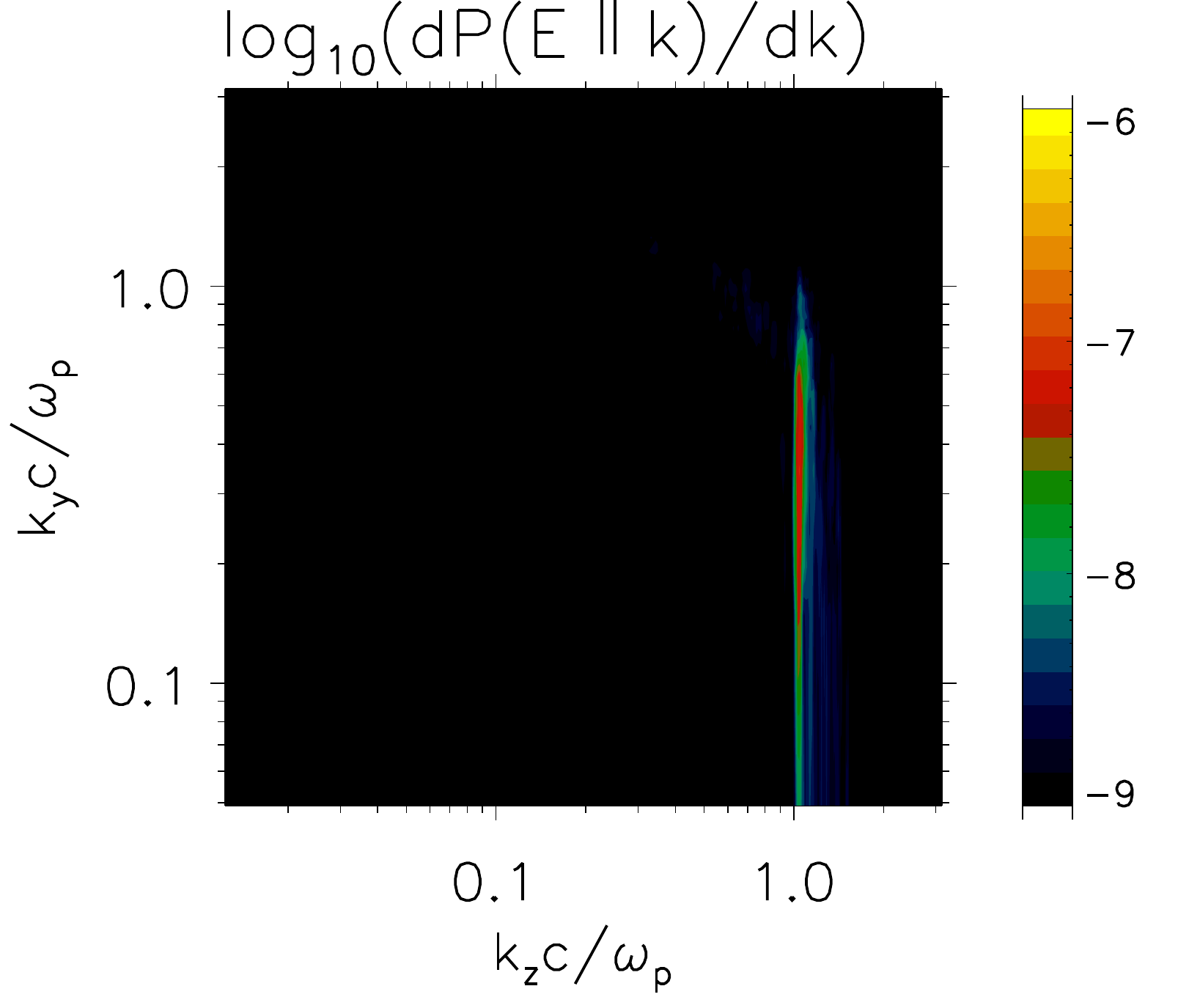}
   \label{eparg549} 
}\quad
\subfloat[ ]{
   \includegraphics[width=0.99\columnwidth]{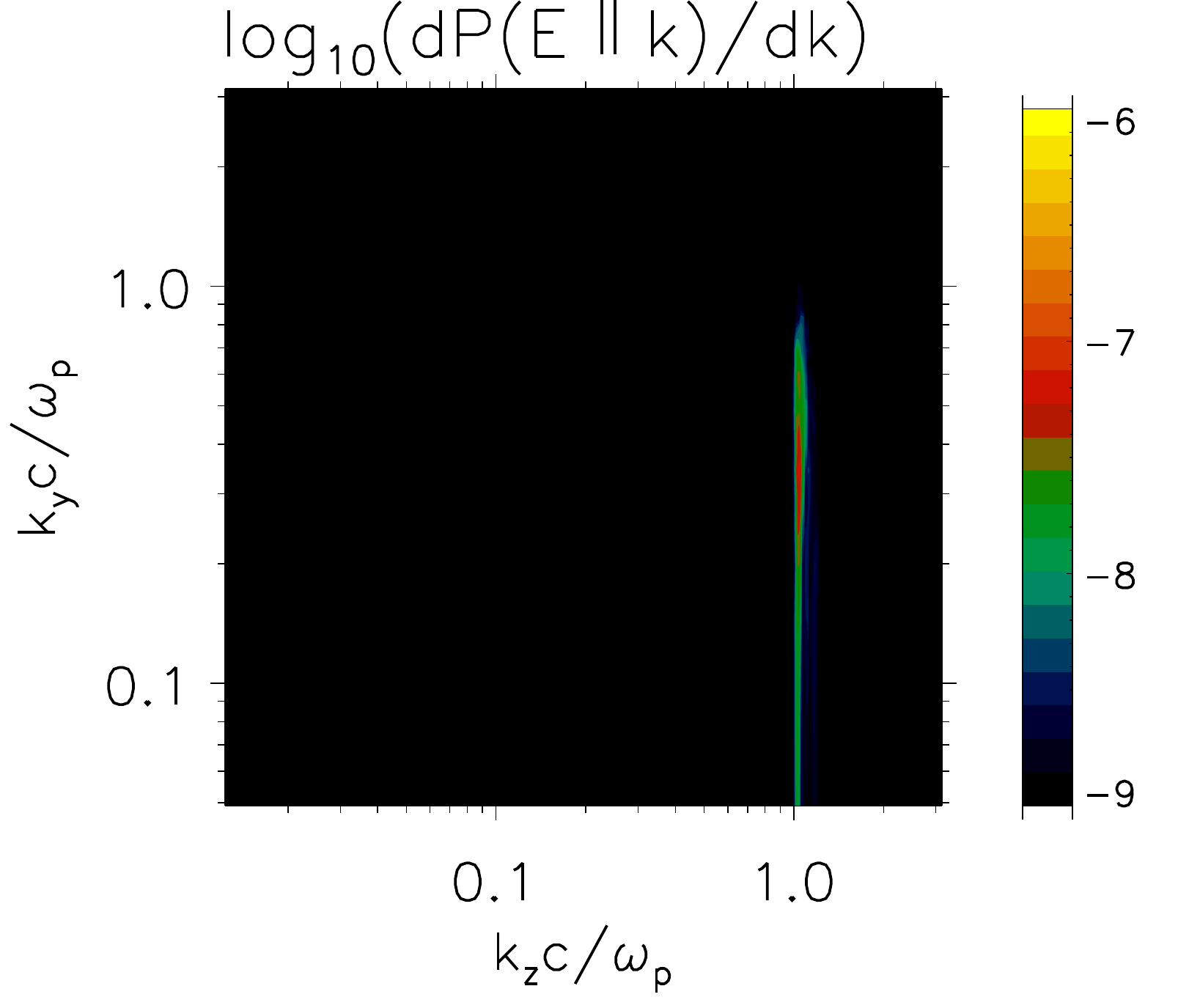}
   \label{epart318}
}
\caption{(a) Two-dimensional Fourier spectrum of $\mathbf{E} \parallel \mathbf{k}$ (in units $(\omega_{p,e}cm_e/e)^2$) at $\omega_{p}t\approx10152$ for run 1. (b) The same as (a), but for run 2.}
\end{figure}

Figs. \ref{eparg549} and \ref{epart318} illustrate the Fourier spectrum of the parallel electric field for runs 1 and 2, respectively, at the same 
time, $\omega_pt\approx10152$. The dimensionless Fourier amplitude is defined as
\begin{multline}
\tilde E(k_{m,n})=\frac{e}{m_e c \omega_p} \frac{1}{N_z N_y} \sum_{i=0}^{N_z-1} \sum_{j=0}^{N_y-1} \\
 \mathbf{e}_k\cdot \mathbf{E}(z_i,y_j)\, \exp\left(2\pi\,\imath
\left[\frac{m i}{N_z}+\frac{n j}{N_y}\right]\right)\ ,
\label{eq_fourier}
\end{multline}
where $m$ and $n$ are the index numbers of the wave-vector components in $z$ and $y$ direction, respectively, and 
$\mathbf{e}_k$ is the unit wave vector. With this definition the electrostatic field energy density, $U_\mathrm{ES}$, is
\begin{equation}
U_\mathrm{ES}={n_e\,m_e c^2\over 2}\sum_{m=0}^{N_z-1}  \sum_{n=0}^{N_y-1}\,\vert\tilde E(k_{m,n})\vert^2.
\label{estat}
\end{equation}

The peak growth of the electric field is observed as expected in quasi-parallel direction to the beam (in agreement with the linear growth rate, Fig. \ref{muR_2_5_g1_5_0_9__g2_20_0_1__Calc_no_sing_calc2}), and the wave intensity in run 1 is also a bit higher than in run 2. In total, the wave energy density represents only a fraction $\lesssim 10^{-2}$ of the beam energy, and the beam does not suffer a significant loss of energy.

\subsection{Non-linear instability saturation}
We now demonstrate that in our simulations the electrostatic instability is not affected by non-linear Landau (NL) damping.  Its rate can be calculated as \citep{Breizman72}  
\begin{multline} 
\omega_{NL}= {3(2\pi)^{1/2}\over 64} \int d^3 k' {W({\bf k}')\over n_e m_e u_i} 
{ ({\bf k}{\bf k}')^2\over (k k')^2} {k^2-k'^2 \over |{\bf k}-{\bf k}'|} \\ \times 
\exp\l[ - {1\over 2} \l( {3u_e^2\over 2 \omega_p u_i} {k^2-k'^2 \over |{\bf k}-{\bf k}'|} \r)^2\r],
\label{nld1}
\end{multline}
where $u_{e,i}=(T_\mathrm{IGM}/m_{e,i})^{1/2}$ denotes the thermal velocity of IGM electrons and ions, and $W_k({\bf k})$ is the spectral energy density of the electric field. With the discrete Fourier amplitudes calculated according to Eq.~(\ref{eq_fourier}) the nonlinear damping rate is
\begin{multline} 
\omega_{NL}= \frac{3(2\pi)^{1/2}}{128} \frac{c^2}{u_i} \sum_{m=0}^{N_z-1}  \sum_{n=0}^{N_y-1} \,\vert\tilde E(k_{m,n})\vert^2 \\ \times 
{ ({\bf k}{\bf k}')^2\over (k k')^2} {k^2-k'^2 \over |{\bf k}-{\bf k}'|} 
\exp\l[ - {1\over 2} \l( {3u_e^2\over 2 \omega_p u_i} {k^2-k'^2 \over |{\bf k}-{\bf k}'|} \r)^2\r].
\label{nld1a}
\end{multline}
The range of wave vectors in which NL damping is efficient, is determined by the Gaussian function in the integrand of Eq. (\ref{nld1}). Therefore, we are interested in wave vectors for which the argument of the e-function is below unity,
\be 
\frac{\vert k^2-k'^2\vert}{|{\bf k}-{\bf k}'|}< \xi {\omega_p\over c},
\label{nld2}
\ee
where $\xi=(2\sqrt{2}/3)(m_e/m_i)^{1/2}(m_e c^2/T_\mathrm{IGM})^{1/2}$. In our simulations $kT_\mathrm{IGM}= 2$~keV and hence $\xi\approx 0.22$. As the growth rate is sharply peaked near $k_z\simeq \pm \omega_p/c$, and we also have $P_k (-\mathbf{k})=P_k (\mathbf{k})$ for the real-valued electric field, we find from Eq. (\ref{nld2}) that NL scattering is essentially always efficient for ${k_z}'\simeq - {k_z}$ and conditionally efficient for ${k_z}'\simeq {k_z}$, if $k_y\approx -k_y'$. Figures~\ref{eparg549} and \ref{epart318} indicate a high wave intensity in a stretch $0.2\lesssim k_y c/\omega_p \lesssim 0.5$, which is well resolved with the $\Delta k_y c/\omega_p \simeq 0.05$ afforded by the simulation grid. The grid resolution in beam direction is chosen sufficiently high to resolve the resonance in the first place, $\Delta k_z c/\omega_p \simeq 0.0123$. Around $\omega_p t\approx 10,000$ we observe a high intensity extending over at least five grid points (in $k$), and the peak value is less than 50\% of the total intensity at these five grid points. We conclude that the grid resolves the intensity at the resonance sufficiently well to permit NL damping in the simulation. 
Using the simulation data, we obtain at the wavevector of peak growth $\omega_{NL}\approx 10^{-5}\omega_p$, which is much smaller than the maximum linear growth rate $\omega_i\approx10^{-3}\omega_p$, and we conclude that NL damping is not responsible for the saturation of the electrostatic instability in our simulations.

Another crucial stabilization mechanism of the electrostatic mode is the modulation instability \citep{Papadopoulos75,Galeev77}. Under the condition 
\be 
{U_{ES}\over n_eT_e}> \max\l[ {\Delta k\over k} (k\lambda_D)^2,\text{ }{m_e\over m_p} \r]
\label{mi1}
\ee
the growth rate of the modulation instability can be found as \citep{Papadopoulos75} 
\be 
\omega_M=\omega_p\l( {m_e\over m_p} {U_{ES}\over n_e T_e} \r)^{1/2}.
\label{mi2}
\ee
Here, $\Delta k$ is the characteristic width of the electric field spectrum. For runs 1 and 2 (see Fig. \ref{muR_2_5_g1_5_0_9__g2_20_0_1__Calc_no_sing_calc2}), the saturation energy of the electric field is $U_{ES}\approx 10^{-19}$ J/m$^3$. The IGM plasma density in the simulations is $n_e=0.25$ m$^{-3}$ which yields $U_{ES}/(n_eT_e)\approx 5\cdot10^{-3}$. Since the electric field energy is localized near the wave number $k\approx\omega_p/c$, $(k\lambda_D)^2\approx 4\cdot10^{-3}$, and condition (\ref{mi1}) is fulfilled for both simulation runs. The damping rate  is  $\omega_M\approx1.6\cdot 10^{-3}\omega_p>\omega_i$, and so it is the modulation instability that stabilizes the electrostatic mode in our simulations. 

Having established that the electrostatic instability has a short growth time compared to other timescales of interactions of the pair beam, we now have to estimate the saturation level and energy transfer rate. In our PIC simulations the modulation instability saturates the waves at an energy density that corresponds to 1\% of the energy density of the beam. Saturation does not imply a vanishing energy transfer, and in fact the beam interacts with the wavefield as long as it does not cool away by inverse-Compton scattering or is deflected out of resonance by ambient magnetic field. Following the beam's propagation through the saturated wave field for the mean free path for Compton scattering, $\lambda_\mathrm{IC}$, is impossible with PIC simulations, and so we resort to analytical estimates.

Saturation is defined as statistical balance between driving, cascading, and damping of waves. Writing the driving rate in the Fourier power as $2\omega_i\,\vert P_k\vert^2$, one would use Parseval's theorem to calculate the energy transfer rate per volume $V$, accounting for equipartition between the kinetic energy and electrostatic energy in the electrostatic mode.
\begin{equation}
\frac{dU_\mathrm{beam}}{dt}=-2\frac{dU_\mathrm{ES}}{dt}=-\frac{2\,\epsilon_0}{V}\int d\mathbf{k}\ \vert P_k\vert^2\,\omega_i\ . 
\label{wave-eq}
\end{equation}
The inverse loss time scale for beam energy then is
\begin{equation}
\tau_\mathrm{loss}^{-1} = \bigg\vert \frac{d\ln U_\mathrm{beam}}{dt}\bigg\vert\simeq \frac{2\,\epsilon_0}{V\,U_\mathrm{beam}}\int d\mathbf{k}\ \vert P_k\vert^2\,\omega_i\ .
\label{wave-eq1}
\end{equation}
Written for the 2D discrete-Fourier power available from our simulation, Eq.~(\ref{wave-eq}) assumes the form
\begin{equation}
\frac{dU_\mathrm{beam}}{dt}=-2\,n_e\,m_e c^2\,\sum_{m=0}^{N_z-1}  \sum_{n=0}^{N_y-1}\,\vert\tilde E(k_{m,n})\vert^2\,\omega_i (m,n)\ ,
\label{wave-eq2}
\end{equation}
and the inverse loss time is 
\begin{equation}
\tau_\mathrm{loss}^{-1} \simeq \frac{2\,n_e\,m_e c^2}{U_\mathrm{beam}}\,\sum_{m=0}^{N_z-1}  \sum_{n=0}^{N_y-1}\,\vert\tilde E(k_{m,n})\vert^2\,\omega_i (m,n)\ .
\label{eq-losstime}
\end{equation}
The symmetry in $P_k$ permits performing the sum over only one quadrant of $\mathbf{k}$ space and then applying a factor $4$ plus another small factor that accounts for extending the wave spectrum from the 2D simulation behavior to the 3D real-world geometry. The energy transfer rate is hence estimated by folding the wave power spectrum close to the saturation level with the linear growth rate. This is a conservative estimate, because already at moderately nonlinear amplitudes the growth rate is observed to fall below 70\% of its initial value. 

At time $t\approx10^4/\omega_p$, our numerical results provide $dU_\mathrm{beam}/dt\approx 6\cdot10^{-21}$ J/(m$^3\cdot$s) and $\tau_\mathrm{loss}\approx 5\cdot10^5/\omega_p$. The wave intensity remains near the peak value for a time period of $\lesssim 5\cdot10^3/\omega_{p}$ which is much shorter than the estimated loss time. The estimated energy loss is below per-cent level, and indeed only a tiny beam energy loss can be observed in the simulations. Numerically, we would have obtained the same result, had we replaced $\omega_i (m,n)$ in Eq. (\ref{wave-eq}) with half of its peak value,
\be
\frac{dU_\mathrm{beam}}{dt}\simeq -2\, U_\mathrm{ES}\,\omega_{i, \mathrm{max}}\ .
\label{loss-scale}
\ee
We shall use this formula when estimating the loss rate of realistic beams. 


\section{Non-linear instability saturation for realistic blazar-induced pair beams}
We shall now discuss the saturation process and level for realistic beam parameters. We begin with the modulation instability as the damping process that we found to dominate in our simulation.
Let us introduce additional scalings: $\gamma=10^6\gamma_6$, $T_\mathrm{IGM}=10^{4}\,T_4$ K. 
Then, the peak linear growth rate of the electrostatic mode can be written as 
\be
\omega_{i,\mathrm{max}}\simeq \left(1.88\cdot 10^{-7}\right)\,\omega_p\,\frac{n_{b20}}{n_{e7}}=5.64\cdot10^{-9}\omega_{p}\ ,
\label{es-rea-gr}
\ee
where in the last expression we inserted $n_{b20}=0.03$ as for the pair distribution shown in Fig.~\ref{Pairspectrum}. Note that the estimate (\ref{es-rea-gr}) is by factor 10 higher than that in \citet{Miniati13} who additionally considered IC-cooled pair beams. We shall now estimate the wave intensity for which the modulation instability is strong enough to halt further wave growth. 
Eqs. (\ref{mi1})-(\ref{mi2}) read
\begin{align} 
\delta&={U_\mathrm{ES}\over n_b\,\gamma_b\, m_ec^2} \nonumber \\
&>\max\l[4\cdot10^{-5}\, {\Delta k\over k} {n_{e7}\,T_4^2\over n_{b20}\,\gamma_6},\text{ }0.1
{ n_{e7}T_4\over n_{b20}\,\gamma_6 }\r] ,
\label{nls1}
\end{align}
\be 
\omega_{M}/\omega_p= 5.2\cdot10^{-3} \sqrt{\delta\, n_{b20}\,\gamma_6\over n_{e7}\, T_4}.
\label{nls2}
\ee

Setting $\omega_M=\omega_{i,\mathrm{max}}$ would require $\delta \approx 10^{-11}$, for which condition (\ref{nls1}) is not fulfilled, and the result of \citet{Papadopoulos75} does not apply. Instead, since $\omega_M\ll kv_{T,i}$, where $v_{T,i}=\sqrt{T_\mathrm{IGM}/m_i}$, we should use the growth rate and threshold condition of the modulation instability Eqs. (10)-(11) derived by \citet{Baikov77}. We write them, respectively, in the form
\be 
\omega_M= -\Gamma_L +|\Delta| \l( -1 -0.25{U_{ES}\over n_eT_e} {\omega_{p}\over\Delta} \r)^{1/2},
\label{nls3.1}
\ee
\be
\delta_M=\frac{U_\mathrm{ES}}{n_b\,\gamma_b\, m_ec^2}> \delta_\mathrm{min}=\frac{4\,n_e T_e}{n_b\,\gamma_b\, m_ec^2} \,{\Gamma_L^2+\Delta^2\over |\Delta|\omega_{p}}
\label{nls4}
\ee
Here, $\Gamma_L$ is the damping rate due to particle collisions and linear Landau damping, whereas $\Delta$ is the mismatch between the Langmuir frequency and the frequency of the unstable oscillations. In the cold-beam limit, it is easy to find \citep{Baikov77}:
\be
\Delta\approx -{\omega_{p} \over 2^{4/3}} \l(n_b\over n_e\gamma \r)^{1/3} \l( \sin^2\theta + {\cos^2\theta\over \gamma^2} \r)^{1/3},
\label{nls5}
\ee
where $\theta$ is the angle between wave vector and the beam. For the realistic beam, the strongest mode develops at the angle $\theta\approx 0.33$ rad (see Fig. \ref{Spread_1overGamma_ksi2_calc2}), yielding for typical parameters 
\be 
\Delta\approx -8.8\cdot10^{-8}\omega_p \l( n_{b20}\over n_{e7} \gamma_6 \r)^{1/3}.
\label{delta}
\ee
Since the collision frequency $\nu_{ei}\approx 10^{-13}\omega_p$, we have $\Gamma_L\ll\Delta$, and the threshold condition for the modulation instability (\ref{nls4}) finally becomes

\be 
\delta_\mathrm{min}\simeq 6\cdot10^{-6}\ T_4\, \gamma_6^{-4/3} 
\,\left(\frac{n_{e7}}{n_{b20}}\right)^{2/3}\simeq 10^{-5}\ ,
\label{nls6}
\ee
where again the last expression applies for the parameters of the pair distribution shown in Fig. \ref{Pairspectrum}, $n_{b20}=0.03$ and $\gamma_6=4$.
Now we solve equation~\ref{nls3.1} for $\omega_M=\omega_{i,\mathrm{max}}$ and find
\be 
\delta_M \simeq \delta_\mathrm{min}\l[1+ 4.7\,\gamma_6^{2/3}\,\left(\frac{n_{b20}}{n_{e7}}\right)^{4/3}\r] \simeq 10^{-5}\ . 
\label{nls7}
\ee
To be noted from this expression is that while formally $\delta_M$ is larger than the threshold value, it is numerically very similar for the parameters of realistic pair beams.

Turning to NL damping, it can be easily seen that the Gaussian in Eq.~(\ref{nld1}) essentially always returns unity, since the cut-off scale $\xi\approx 14$ (cf. Eq.~(\ref{nld2})). Most electrostatic energy grows near the resonance wave number $\omega_p/c$ (see Fig. \ref{Spread_1overGamma_ksi2_calc2}), and so we can analytically estimate the second and third factor of the integrand in Eq.~(\ref{nld1}) and find them approximately equal to $k-k'$. Writing $\Delta k =k-k'\approx 0.1\,\omega_p/c$, we obtain
\begin{align} 
\omega_\mathrm{NL}&\approx 10^{-3}\ \omega_p\, \l( \Delta k\ c\over \omega_p \,\r) {\delta\, n_{b20}\,\gamma_6\over n_{e7}\, T_4^{1/2} } \nonumber \\
 & \approx 10^{-4}\  \omega_p\, \frac{\delta\, n_{b20}\, \gamma_6}{ n_{e7}\, T_4^{1/2} }.
\label{nls3}
\end{align}
The estimate for $\Delta k$ is very rough, but our result only weakly depends on this parameter, as will be seen below. 
For the pair distribution shown in Fig. \ref{Pairspectrum}, $n_{b20}=0.03$, $\gamma_6=4$, setting $\omega_{NL}=\omega_{i,\mathrm{max}}$ leads to 
\be 
\delta_{NL}\approx 5\cdot10^{-4} . 
\label{nls8}
\ee
This corresponds to a higher wave intensity as that found for the modulation instability (cf. \ref{nls7}), and we conclude that the modulation instability provides a stronger limitation on the beam intensity than does NL damping. Note that \citet{RS12} and \citet{Miniati13} used Eq. (\ref{mi1}) as the threshold condition for the modulation instability, which we show to be not applicable for the realistic parameters.

Using our result for $\delta_M$, we can estimate the relaxation time of the blazar-induced pair beam using Eq.~(\ref{loss-scale}),
\begin{align}
\tau_\mathrm{loss}^{-1} &\simeq 2\,\delta_M \,\omega_{i,\mathrm{max}} \nonumber \\
&\simeq (6\cdot 10^{-11}\ \mathrm{s}^{-1})\,T_4\,\gamma_6^{-4/3} 
\,\left(\frac{n_{b20}}{n_{e7}}\right)^{1/3} \ .
\label{real-loss}
\end{align}
This loss rate is to be compared with that for inverse-Compton scattering at redshift $z$,
\be
\tau_\mathrm{IC}^{-1} \simeq (1.3\cdot 10^{-14}\ \mathrm{s}^{-1})\,\gamma_6\,(1+z)^4\ .
\label{loss-IC}
\ee
For the ratio of timescales we find with the realistic parameters $n_{b20}=0.03$ and $\gamma_6=4$, and introducing the redshift scaling for the IGM density
\begin{align}
\frac{\tau_\mathrm{loss}}{\tau_\mathrm{IC}} &\simeq 2.2\cdot 10^{-4}\ \frac{\gamma_6^{7/3}}{T_4}\,
\left(\frac{n_{e7}}{n_{b20}}\right)^{1/3}\,(1+z)^5 \nonumber \\
&\simeq 0.02\,(1+z)^5\ , 
\label{loss-final}
\end{align}
indicating that for redshift $z\lesssim 1.2$ plasma instabilities drain the energy of the pair beam faster than comptonization of the microwave background would. This estimate is very weakly dependent on the density of the pair beam and that of the IGM, and so the distance from the AGN is of moderate importance, as is the TeV-band luminosity. It does strongly vary with the choice of pair Lorentz factor. With our nominal $\gamma_6=4$ the pairs would up-scatter the microwave background to about 10~GeV in gamma-ray energy, i.e. into the energy range where the \textit{Fermi}-LAT has the optimal sensitivity for the cascade signal. A number of comments are in order:
\begin{itemize}
\item In our simulations the peak growth rate was somewhat reduced during the nonlinear phase which would imply a longer loss time, if TeV-scale pair beams behaved in the same way.
\item Any other cascading or loss mechanism beyond the modulation and the non-linear Landau damping considered here would also increase the loss time, because it would reduce the saturation amplitude, $\delta_M$.
\item Substantial uncertainty derives from the exact form of the pair spectrum, and \citet{Miniati13} find a peak growth rate around 10\% of our result only by allowing for efficient cooling, which would translate to a ten times longer loss time. Typically, the exact primary gamma-ray spectrum is not known, in particular neither the spectral index nor the cut-off energy, and so it is not possible to completely predict the pair spectrum.
\item We calculate the bulk energy loss of the pair beam without consideration of its energy dependence. It is possible that the energy loss primarily affects the pairs that are also instrumental in driving the electrostatic mode, while leaving unaffected those pairs that are most efficient in producing the gamma-ray cascade signal in the GeV band.
\end{itemize}
All the above suggests that the result~(\ref{loss-final}) should be seen as lower limit with significant uncertainty.
Thus, we conclude that the blazar-induced pair beam at moderate redshift $z\approx 0.2$ will lose its energy with similar efficiency to inverse-Compton scattering and to the interaction with plasma waves.

\section{Summary}\label{Summary}

We revisited the growth and feedback of the electrostatic instability under conditions relevant for blazar-induced pair beams propagating through the IGM. Our goal was to clarify somewhat contradicting statements by \citet{RS13} and \citet{Miniati13} and to establish the energy loss rate of pair beams for driving the waves. 

First of all, we calculated the energy distribution function of blazar-induced pairs without modification by IC scattering, which is appropriate if the electrostatic instability provides the dominant energy loss. We assumed a power-law spectrum $\propto E^{-1.8}$ for the blazar emission and used the models of \citet{Finke10} and \citet{Fabian92} to describe the EBL spectrum. We found a broad pair spectrum ($10^1<\gamma<10^8$) similar to that in \citet{Miniati13}. This spectrum contains a pronounced low-energy bump arising from interaction with the X-ray background that is missing in the treatment of \citet{RS12a,RS12,RS13}. Then, we used the newly evaluated pair distribution to study the growth rate of the electrostatic instability and did not find any significant effect of low-energy pairs with $\gamma<10^4$, lending support to the results of \citet{RS12,RS13}.

As the absence or presence of cascade emission at a few GeV is the most important observable, we have to consider the pair spectrum at a distance from the blazar where the bulk of the pairs is produced that re-radiate into this energy band, which is about 50~Mpc. The accumulation of pairs is limited to a much shorter pathlength that does not exceed the loss length to inverse-Compton scattering of the microwave background, for which we use that of pairs with Lorentz factor $\gamma=10^7$.

We investigated the growth rate of the instability for arbitrary wave vectors considering in particular the effect of the transverse beam temperature (realistics beams have $\mathrm{rms}(p_\perp)\approx m_ec/2$). If the beam had no angular spread, then the growth rate would reach its maximum at wave vectors perpendicular to the beam. However, for a realistic finite angular spread the growth rate is the largest at wave vectors quasi-parallel to the beam direction, and the maximum growth rate is reduced, in agreement with \citet{Miniati13}, but it is still by more than a factor of ten larger than the peak growth rate of the strictly parallel electrostatic mode studied by \citet{RS13}.

\citet{Miniati13} assumed IC cooling of all beam electrons and positrons, not only those with $\gamma > 10^7$. Their results indicate that the IC scattering reduces the maximum growth rate by about an order of magnitude. As we investigate the viability of the instability providing the dominant energy loss, we need to 
consider an uncooled beam, and hence the larger linear growth rate. The strong dependence of the growth rate on the shape of the pair spectrum suggests that there may also be substantial variation in the growth rate arising from the shape of the primary gamma-ray spectrum produced by the blazar, in particular the spectral index and the cut-off energy. 


We then studied the non-linear beam evolution with PIC simulations. In contrast to earlier studies \citep{Sironi14,Kempf16,Rafighi17}, we did not consider a relativistic Maxwellian for the beam, but superposed relativistic Maxwellians to mimic a realistic beam spectrum, albeit at small beam Lorentz factor following \citet{Rafighi17}. In the simulation the beam loses about 1\% of its initial energy, and the saturation level of the electrostatic waves is determined by the modulation instability. Analytical analysis suggests that beam relaxation should be achieved on the timescale  $\approx 5.6\cdot10^5/\omega_p$, much longer than the time covered by our simulation ($\approx 1.5\cdot10^4/\omega_p$).

Our analytical analysis then permits extrapolation to realistic pair beams. We determine the linear growth rate of the electrostatic instability and find that also in this case the modulation instability is a faster saturation process than is nonlinear Landau damping. \citet{Miniati13} arrived at the opposite conclusion, but used a less accurate threshold condition of the modulation instability. Balance of growth and damping determines that saturation level, from which we analytically estimate that the energy-loss time scale for beam instabilities is slightly smaller than that for comptonization of the microwave background, so that the electrostatic beam stability could at least reduce the intensity of the gamma-ray cascade emission in the GeV band. The uncertainties in the estimate are large though, and there is a significant dependence on redshift.

If the effective loss length is indeed slightly reduced by beam instabilities, the flux of the cascade signal is correspondingly smaller. Any magnetic deflection of the beam would then have to be accomplished over this smaller pathlength and would hence require a stronger magnetic field. An interesting possibility is that the intergalactic magnetic field increases the transverse momentum spread and hence reduces the growth rate of the electrostatic instability. A strictly homogeneous magnetic field would also diverge the electron beam and the positron beam, which might trigger other plasma instabilities. A detailed study of these effects is beyond the scope of this paper. 


\acknowledgments

The numerical simulations were performed with the EPOCH code that was in part funded by the UK EPSRC grants EP/G054950/1, EP/G056803/1, EP/G055165/1 and EP/ M022463/1. The numerical work was conducted on resources provided by The North-German Supercomputing Alliance (HLRN) under project bbp00003. M.P. acknowledges support through grant PO 1508/1-2 of the Deutsche Forschungsgemeinschaft. The work of J.N. is supported by Narodowe Centrum Nauki through research project DEC-2013/10/E/ST9/00662.

\appendix
\section{Analytical approximation for stellar radiation spectrum at redshift 0.2}

We approximate the energy spectrum of stellar radiation at redshift $z=0.2$ by 
\be 
f(\epsilon)= \sum_{i=1}^4 {N_i\over \Gamma(1+q_i)k_B T_i}\l( \epsilon\over k_B T_i \r)^{q_i}\exp\l( -{\epsilon\over k_B T_i}\r), 
\label{app1}
\ee
where $\epsilon$ is in eV, $f(\epsilon)$ in eV$^{-1}$ cm$^{-3}$. Other parameters are listed in Table \ref{Tabfit}.
\begin{table}[h]
\caption{Fitting parameters for the approximation Eq. (\ref{app1})} 
\centering 
\begin{tabular}{|c|c|c|c|} 
\hline 
   $i$     & $q_i$ & $k_BT_i$, eV & $N_i$, cm$^{-3}$ \\ [0.5ex]
\hline  
1 & 0.5 & $3.3\times10^{-3}$ & 0.78  \\
\hline
$2$ & 0 & 0.53 & 0.01 \\
\hline
$3$ & 0 & 0.04 & 0.035  \\
\hline
$4$ & 0 & 2 & 0.0007  \\ [1ex]
\hline
\end{tabular}
\label{Tabfit} 
\end{table}

\section{Analytical approximation for the pair distribution function}

The pair distribution function can be approximated as
\begin{multline} 
{f_b(\gamma)\over n_b}= N_1 \l( \gamma\over \gamma_1\r)^{-b_1}\exp\l(-\sqrt{\gamma_1\over\gamma}\r)\Theta[(\gamma-6\times10^3)(10^{8}-\gamma)] \\
+ N_2\l( \gamma\over \gamma_2\r)^{b_2}\exp\l( - \l( \gamma\over\gamma_2\r)^{0.7}\r) 
+ N_3\l( \gamma\over \gamma_3\r)^{b_3}\exp\l( - { \gamma\over\gamma_3} \r) , 
\label{app2}
\end{multline}
where the fitting parameters are summarized in Table~\ref{Tabfit2}.

The approximation used by \citet{RS12a} is
\be
\frac{f_b(\gamma)}{n_b}\approx \frac{\gamma^{1/2-\alpha}}{\gamma_c^{3/2-\alpha}\Gamma(\alpha-3/2)} \frac{\exp(-\gamma_c/\gamma)}{1+(\gamma/\gamma_b)^{3/2}}, 
\label{rdf10}
\ee
where $\gamma_c=M_c/\ln\tau_0$, $\gamma_b=M_c\tau_0^{2/3}/2^{7/3}$, $\tau_0=10^3$, $M_c=2\times10^6$, $\alpha=1.8$.

\begin{table}[h]
\caption{Fitting parameters for the approximation Eq. (\ref{app2})} 
\centering 
\begin{tabular}{|c|c|c|c|} 
\hline 
   $i$     & $b_i$ & $\gamma_i$ & $N_i$ \\ [0.5ex]
\hline  
1 & 1.6 & $10^{6.2}$ & $3\times10^{-7}$  \\
\hline
$2$ & 1.8 & $10^{2.2}$ & $1.1\times10^{-7}$ \\
\hline
$3$ & 1.8 & $10^{3.2}$ & $1.8\times10^{-8}$  \\ [1ex]
\hline
\end{tabular}
\label{Tabfit2} 
\end{table}


\end{document}